\RequirePackage[2020-02-02]{latexrelease}
\documentclass[arXiv,onecolumn,
superscriptaddress]{revtex4}
\usepackage[intlimits]{amsmath}
\usepackage{amsfonts}
\usepackage{graphics}
\usepackage{bbm}
\usepackage{subfigure}
\usepackage[usenames]{color}
\usepackage{epstopdf,epsfig}
\usepackage{mathtools}
\usepackage{ulem}
\usepackage{soul}
\usepackage{bm}
\usepackage{float}

\DeclarePairedDelimiterX\braket[2]{\langle}{\rangle}{#1 \delimsize\vert #2}

\usepackage{indentfirst}

\usepackage{graphicx}
\usepackage{amsmath}
\usepackage{amsfonts}
\newcommand\be            {\begin{equation}}
\newcommand\bea           {\begin{equation}\begin{array}l\displaystyle}
\newcommand\ee            {\end{equation}}
\newcommand\bes           {\begin{subequations}}
\newcommand\esu           {\end{subequations}}

\makeatletter
\newcommand{\vast}{\bBigg@{3.4}}
\newcommand{\Vast}{\bBigg@{5}}

\makeatother

\def\3pt#1#2#3{{\langle{#1}\vert{#2}\vert{#3}\rangle}}
\def\barray{\begin{eqnarray}}
\def\earray{\end{eqnarray}}

\usepackage{tikz}
\usetikzlibrary{calc}

\begin{document}

\title{Parametrizations of the solutions of Love equations and their momenta
}

\author{Marco Faccioli\footnote{Present address: Scuola Campostrini, Via S. Maria in Organo 2, I-37100, Verona, Italy}}
\affiliation{CNR-IOM DEMOCRITOS Simulation Center, Via Bonomea 265, I-34136 Trieste, Italy} 
\author{Andrea Colcelli\footnote{Present address: I.I.S. Campus ``Leonardo da Vinci", Via Tusicum, I-06019, Umbertide, Italy}}
\affiliation{SISSA and INFN, Sezione di Trieste, Via Bonomea 265, I-34136 
Trieste, Italy}
\author{Danny Laghi}
\affiliation{Physik-Institut, Universit\"at Z\"urich, Winterthurerstrasse 190, 8057 Z\"urich, Switzerland}
\affiliation{CNR-IOM DEMOCRITOS Simulation Center, Via Bonomea 265, I-34136 Trieste, Italy} 
\author{Yu Zhou}
\affiliation{School of Science, Jiangsu University of Science and Technology, Jiangsu 212100, China}
\author{Giacomo Gori}
\affiliation{CNR-INO, Area Science Park, Basovizza, I-34149 Trieste, Italy}
\author{Andrea Trombettoni}
\affiliation{Department of Physics, University of Trieste, Strada Costiera 11, I-34151 Trieste, Italy}
\affiliation{CNR-IOM DEMOCRITOS Simulation Center, Via Bonomea 265, I-34136 Trieste, Italy} 

\begin{abstract}
  The Love equation is a paradigmatic example of a Fredholm integral equation
  of great interest in mathematics and physics, where its applications range
  from electrostatic problems to exactly solvable
  models for one--dimensional systems. In this work we propose a method to approximate
  the normalized solutions of Love equations of the first and second kind
  via a parametrization obtained by
  Hermite interpolation at Chebyshev nodes. Following the same approach, we also obtain
  approximations for the second, fourth and sixth momenta of the solution of the Love equations.
  We focus in particular on one--dimensional quantum gases described by the Lieb--Liniger and the
  Gaudin--Yang models. For these models, the solutions of Love equations correspond to the
  normalized densities of pseudorapidities. We show that remarkably accurate
  results can be obtained using the Hermite--Chebyshev interpolation with
  a number of points as low as $17+2$.
  In the case of the Lieb--Liniger model, we also propose
  a very accurate closed--form estimate of the dimensionless ground state energy valid for all values of the dimensionless interaction parameter incorporating
  known results for small and large couplings supplemented by the Hermite--Chebyshev interpolation for intermediate couplings.
  As an application of this result, we compute
  the dimensionless local two-particle correlation function and the black soliton solutions
  of the Lieb--Liniger model.
\end{abstract}

\maketitle

\section{Introduction}
The Love integral equation was first studied by E. R. Love to describe the electrostatic field of
a capacitor consisting of two parallel circular plates \cite{Love49}. Since then, its applications
have extended far beyond this original setting \cite{Farina2022}.
For the original problem of a condenser formed by parallel, circular, coaxial disks,
Love wrote the electrostatic potential as:
\begin{equation} \label{potentialLove}
V(\rho, \xi, \xi')  = \frac{V_{0}}{\pi}\int_{-1}^{1} \left\{\frac{1}{\sqrt{\rho^{2} + (\xi + it)^{2}}} \pm \frac{1}{\sqrt{\rho^{2} + (\xi' + it)^{2}}}\right\}f(t)\,dt\,,
\end{equation}
where dimensionless coordinates normalized to the (common) plate radius have been used, and the real part of the integral is understood. In Eq. \eqref{potentialLove} $\rho$ is the radial distance, $\xi$ and $\xi'$ are the axial distances from the two plates, while $V_{0}$ is the magnitude of the potential in each plate. Finally, $f(t)$ is the solution to the Fredholm integral equation:
\begin{equation}\label{originalLove}
f(x) \pm \frac{1}{\pi} \int_{-1}^{1} \frac{\kappa}{\kappa^{2} + (x-t)^{2}}f(t)dt  = 1 \, , \quad -1\leq x \leq 1\,,
\end{equation} 
where $\kappa$ is the spacing parameter, which is equal to the ratio between the distance and the radius of the disks. The upper sign in Eqs. (\ref{potentialLove}) and (\ref{originalLove}) refers to the case of equally charged plates, while the lower one refers to the case of oppositely charged disks. In what follows we shall refer, for convenience, to Eq. (\ref{originalLove}) with the lower (upper) sign as ``Love equation of the first (second) kind".
Eqs. (\ref{originalLove}) are also referred to as ``Love integral equations".

Among the most interesting settings in which Love equation proves useful \cite{Farina2022}
is the study of one-dimensional (1D) dilute quantum gases \cite{Cazalilla2011}. In this context
there are models which are analytically solvable \cite{KorepinBook,MussardoBook} and 
accurately describe the low-energy dynamics of ultracold atoms in 1D waveguides, thus 
providing an ideal playground for the interplay between theory and experiment
\cite{Cazalilla2011,Guan2013}. The Love equation finds application in the study of the
Lieb--Liniger (LL) and Gaudin--Yang (GY) models \cite{Lieb63,Gaudin67,Yang67,GaudinBook}. Such
models are exactly solvable, \textit{i.e.} integrable, and describe respectively
1D bosonic and fermionic gases with a contact interaction \cite{Cazalilla2011}.
In these two models the many--body eigenfunctions can be found using the Bethe ansatz formalism
\cite{KorepinBook,FranchiniBook} and are expressed in terms of quantities called pseudorapidities,
which can be determined once the boundary conditions are fixed \cite{Lieb63}.

The LL model in itself has a very wide range of applications. For instance, it can be used
in the study of few-body systems \cite{Muga-Snider,Batcherlor-Guan-Oelkers-Lee,Sakmann-Streltsov,Sykes-Drummond} and quench dynamics
\cite{Kormos-Sashi-Chou,Cartarius-Kawasaki-Minguzzi,Caux-Essler,Carleo-Celovani-Sanchez}. The model can moreover be extended to anyonic statistics \cite{Cheon-Shigehara} and supersymmetry \cite{Lundholm2013,Guilarte2017}, and it can be mapped onto a variety of other models, such as Yang-Mills theory on the $2$-sphere \cite{Flassig2016}, three-dimensional black holes \cite{Panchenko}, Kardar-Parisi-Zhang (KPZ) model \cite{Calabrese2014} and directed polymers \cite{DeLuca2015}. The LL model can be viewed as a quantum non-linear Schr\"odinger equation \cite{KorepinBook}. Reviews of experimental investigations of the LL model can be found in \cite{Bloch2008,Jiang2015}, while for experimental works on the GY model we refer to \cite{Moritz2005,Liao2010,Guan2013}. Both models can be mapped to a BCS model \cite{Fuchs2004,Batchelor-Guan-McGuire,Iida2005} and have applications in the study of quantum criticality in ultracold atoms \cite{Zhao2009,Guan2010,Guan2011,Yin2011a}.

As we will review in the following Sections, Eqs. (\ref{originalLove}) have the same form as the equations describing
the densities of pseudorapidities in the GY and LL models. In particular, for the LL (GY)
model the equation for the densities has the same form as Love equation of the first (second) kind.
This fascinating parallelism is of considerable interest. Once the solution of Love equation is known for the LL and GY models, other important physical properties of these systems can be extracted from it. Indeed, knowledge of the density (or weight) function $f(x)$ and of its momenta provides access to correlation functions \cite{Olshanii2003,Cheianov2006,Ristivojevic2014,Lang-Minguzzi,Bastianello2018} and to information on solutions of the one--dimensional
quantum non-linear Schr\"odinger equation \cite{Dunjko2001,Schmidt2006,Peotta2014,Choi2015}, which are the
counterpart of the corresponding stationary and dynamical solutions -- such as
black and grey solitons -- of the
one--dimensional classical non-linear Schr\"odinger equation
\cite{PitaevskiiBook}. An improved mean-field description of the dynamics of the LL model \cite{Peotta2014},
connected to the zero temperature limit of the generalized hydrodynamics approach \cite{Bouchoule2022},
can be obtained by a one-dimensional time-dependent
generalized non-linear Schr\"odinger equation where the non-linear term is not $\propto |\psi|^2$ (where $|\psi|^2$ is the density), but actually a function of $|\psi|^2$ which is determined by the solution of the Love integral equation, see Section \ref{solitons_section}. Therefore to have accurate
parametrized
solutions of the Love equations has considerable practical
importance for the study of the dynamics of the generalized non-linear Schr\"odinger equation description of the LL model.

In this work, we derive numerically approximated solutions to Love equations of the first and second kind and their second, fourth, and sixth momenta. The approximation is constructed by finding a suitable Hermite interpolation formula using Chebyshev polynomials \cite{Fox1968} that reproduces the normalized solution of Love equation while minimizing the error with respect to the numerical solution. Approximate numerical solutions of Love equations have been sought previously in \cite{Piessens1976,Milovanovic2013,Ristivojevic2019}. Our work is characterized by a systematic search for accurate Chebyshev interpolation formulas for the weight functions $f(x)$ solving Love equations of the first and second kind, and for their momenta. After a convenient normalization, we first construct, for each Love integral equation, an interpolation formula with Chebyshev polynomials of the weight functions and we find a maximum relative percentage error with respect to the numerically ``exact" results of about $10^{-6}\%$ and $10^{-3}\%$, for Love equation solution of the first and second kind, respectively. As a first attempt, we use these results to construct a further Hermite--Chebyshev interpolation for the momenta of the solutions. With this approach we observe an accumulation of error in the computation of the momenta, ranging between $10^{-2}\%$ and around $3\%$ for different momenta and/or interactions. We therefore develop a more refined method to interpolate the second momentum of the solution of Love integral equation of the first kind, which is equivalent to the case of the LL gas. In particular the second momentum of the density of pseudorapidities is related to the ground state energy density of the system, and hence a compact and accurate closed form for this function is of interest. Our method consists of dividing the range of the interaction parameter $\gamma$ into three regions: In the first and the last, corresponding to weak and strong coupling, we rely on analytical expressions for the second momentum behaviour, while in the intermediate region we build a Chebyshev interpolation for the energy density function $e(\gamma)$. This procedure yields a simple closed-form analytical expression — to be understood as an accurate numerical approximation — for the energy density with maximum relative percentage error of about $10^{-6}\%$. 

Owing to these very small estimated errors, our expression for $e(\gamma)$ can be used to easily and efficiently compute the local two--particle correlation function and black soliton solutions. These applications are presented in Sections \ref{g_2_section} and \ref{solitons_section} of this work as illustrative examples of our results. 

The paper is organized as follows. In Section \ref{sec:2} we review the electrostatic problem, while in
Section \ref{sec:3} we discuss the connection between the Love integral equations and the
LL and GY models. Section \ref{sec:numsol} is devoted to a discussion
of the numerical solution of these equations. Sections \ref{sec:5} and \ref{sec:6} constitute the core of the paper, where we develop Hermite–Chebyshev parametrizations for the normalized solutions of the Love equations (Section \ref{sec:5}) and their second, fourth, and sixth momenta (Section \ref{sec:6}). In Section \ref{sec:7} we combine analytical results at weak and strong coupling with Hermite–Chebyshev interpolation to obtain an accurate expression for the dimensionless ground state energy of the Lieb–Liniger model over the entire interaction range. We then illustrate the usefulness of this result by computing the local two-particle correlation function in Section \ref{g_2_section} and black-soliton solutions in Section \ref{solitons_section}. Conclusions are presented in Section \ref{concl}.

\section{The electrostatic problem}
\label{sec:2}
For a condenser formed by parallel, circular, coaxial disks of equal radius
held at potentials $V_1$ and $V_2$, the electrostatic potential is described by two distinct weight functions $f_j(t)$, with $j=1,2$,
for the two plates. In particular one has \cite{Felderhof}:
\begin{equation}
\label{Love_eq_1}
f_1(x)+\frac{1}{\pi} \int_{-1}^{1} \frac{\kappa}{\kappa^{2} + (x-t)^{2}}f_2(t)dt  = \frac{2}{\pi}V_1\,,
\end{equation}
for one plate with charge $Q_1 = \int_{0}^{1} f_1(t) dt$, while:
\begin{equation}
\label{Love_eq_2}
f_2(x)+\frac{1}{\pi} \int_{-1}^{1} \frac{\kappa}{\kappa^{2} + (x-t)^{2}}f_1(t)dt  = \frac{2}{\pi}V_2\,,
\end{equation}
for the other plate with charge $Q_2 = \int_{0}^{1} f_2(t) dt$. Writing: $V_2 = \mathcal{Z} V_1$, assuming $\mathcal{Z}>0$, from the above two equations one can write a single Fredholm integral equation for the weight function $\bar{f}(x) = \frac{f_1 + f_2}{2}$ reading
\begin{equation}
\label{two_plates_GY}
\bar{f}(x)+\frac{1}{\pi} \int_{-1}^{1} \frac{\kappa}{\kappa^{2} + (x-t)^{2}}\bar{f}(t)dt  = \frac{2}{\pi} \left(\frac{1+\mathcal{Z}}{2}\right) V_1\,.
\end{equation}
If we instead write: $V_2 = -\mathcal{Z} V_1$, with $\mathcal{Z}>0$, taking: $\tilde{f}(x)=\frac{f_1-f_2}{2}$, we have:
\begin{equation}
\label{two_plates_LL}
\tilde{f}(x)-\frac{1}{\pi} \int_{-1}^{1} \frac{\kappa}{\kappa^{2} + (x-t)^{2}}\tilde{f}(t)dt  = \frac{2}{\pi}\left(\frac{1+\mathcal{Z}}{2}\right)V_1\,.
\end{equation}
Hence, for a condenser whose plates carry charges of equal (opposite) sign but different magnitude, it suffices to compute $f(x)$ from Eq. (\ref{originalLove}) with the plus (respectively, minus) sign and then, after normalizing with $2V_1 /\pi$, multiply the weight function by
$(\mathcal{Z}+1)/2$ to get the solution $\bar{f}(x)$ [$\tilde{f}(x)$]. Finally, when $\mathcal{Z}=1$ in Eq. (\ref{two_plates_GY}), meaning that charges of equal sign have equal magnitude,
one recovers Eq. (\ref{originalLove}) with the plus sign, once $f(x)$ has been normalized with $2 V_1 /\pi$. Likewise,
setting $\mathcal{Z}=1$ in (\ref{two_plates_LL}) gives, as expected, Eq. (\ref{originalLove}) with the minus sign, again after normalizing the weight function as before. 

From the knowledge of the weight function, one can also determine the capacitance of the parallel plate capacitor as \cite{Ristivojevic2020}:
\begin{equation}
\nonumber
C=\frac{1}{\pi} \int_{-1}^1 f(t)\,dt\,.
\end{equation}

As we shall see in detail below, the Love equation of the first (second) kind coincides with the equation governing the density of pseudorapidities of the LL (GY) model. Another interesting case is that of $V_1=V$ and $V_2=0$. In this case, indeed, Eqs. (\ref{Love_eq_1}) and (\ref{Love_eq_2}) describe the densities of pseudorapidities of a one--dimensional Bose--Fermi mixture \cite{Imambekov2006}.
The Love equation of the second kind is also found to describe the density of pseudorapidities and the dressed charge function of a LL gas of anyons \cite{Patu2007}, where the plate separation is replaced by a renormalized coupling constant encoding the statistical properties of the particles.

\section{Lieb--Liniger and Gaudin--Yang models}\label{LLandGY} 
\label{sec:3}
The LL model describes a 1D bosonic gas with repulsive contact interaction
\cite{Lieb63}. In the second quantization formalism the Hamiltonian is given by \cite{KorepinBook}:
\begin{equation}
\label{H_LL}
H_{B} =  \int_{0}^{L}\,dx\, \left[ \frac{\hbar^2}{2m_B} \frac{\partial \Psi^{\dagger}}{\partial x} \frac{\partial \Psi }{\partial x} + \frac{\hbar^2 c_B}{2m_B} \Psi^{\dagger}(x)\Psi^{\dagger}(x)\Psi(x)\Psi(x)\right]\,,
\end{equation} 
where $L$ is the length of the system, $m_B$ the mass of the bosons, $c_{B}$ is the coupling constant and $\Psi(x)$ is the bosonic matter-field. The matter-field satisfies the constraint:
\begin{equation}
\int_{0}^{L} \Psi^{\dagger}(x)\Psi(x)\,dx = N_{B}\,,
\end{equation} 
where $N_{B}$ is the total number (operator) of bosons. To simplify the
notation, we will denote by $N_B$ both the total number of bosons and the total
number operator.
It is convenient and customary
to define a dimensionless interaction parameter as
\cite{Cazalilla2011,PethickBook}:
\begin{equation}
\label{def_gamma_B}
\gamma_{B} \equiv \frac{c_{B}}{n_{B}}\,,
\end{equation} 
where $n_{B}=N_B /L$ is the particle density of the system.

The GY model, in turn, describes a 1D fermionic gas of two spin components with attractive interaction between them. In the second quantization formalism, the Hamiltonian reads \cite{Gaudin67}:
\begin{equation}
H_{F} =  \int_{0}^{L}\,dx\, \left[ \frac{\hbar^2}{2m_F} \left( \frac{\partial \Psi_\downarrow^{\dagger}}{\partial x} \frac{\partial \Psi_\downarrow }{\partial x}+ \frac{\partial \Psi_\uparrow^{\dagger}}{\partial x} \frac{\partial \Psi_\uparrow }{\partial x} \right)+ \frac{\hbar^2 c_F}{2m_F}\Psi_{\downarrow}^{\dagger}(x)\Psi_{\uparrow}^{\dagger}(x)\Psi_{\uparrow}(x)\Psi_{\downarrow}(x)\right]\,,
\end{equation} 
where $m_{F}$ is the mass of the fermions, $c_{F}$ is the coupling constant and $\Psi_{\downarrow}(x)$ and $\Psi_{\uparrow}(x)$ are the fermionic matter-field for the two spin-components. In this case the matter-fields satisfy the constraint:
\begin{equation}
\int_{0}^{L} \left[\Psi_{\downarrow}^{\dagger}(x)\Psi_{\downarrow}(x) + \Psi_{\uparrow}^{\dagger}(x)\Psi_{\uparrow}(x)\right]\,dx= N_{F}\,,
\end{equation}
where $N_{F}$ is the total number of fermions. Note that a contact interaction is possible here only by virtue of the two spin components. As we did for the bosonic system, it is convenient to define a dimensionless interaction parameter denoted by $\gamma_{F}$ and defined as:
\begin{equation}
\gamma_{F} = -\frac{c_{F}}{n_{F}}\,,
\end{equation} 
where $n_{F}=N_F/L$ is the total particle density for the fermionic model. From now on, we will drop the subscripts $B$ and $F$ if the equations are valid for both the LL and GY models.

For the LL model, the Bethe ansatz technique yields an exact expression for the many--body wavefunction having the form \cite{KorepinBook,Guan2013}:
\begin{equation}
  \label{LLpsi} 
 \psi(x_1,\dots,x_N) = \sum_P A(P) e^{i \left(k_{P_1} x_{1}+\dots +k_{P_N} x_{N}\right)}\,, 
\end{equation}
where $P$ runs over all permutations of the integers $\{1,2,\dots,N\}$,
the $A(P)$'s are coefficients determined by the matching conditions 
of the wavefunction, while the $k_j$'s are the pseudorapidities of the system.
An expression similar to (\ref{LLpsi}) holds for the GY model, once the spin degrees of freedom are taken into account. We do not write the explicit expression
of $A(P)$ and refer to textbooks \cite{KorepinBook,GaudinBook} for a detailed discussion of the Bethe ansatz construction, the coefficients
$A(P)$, and the equations determining the pseudorapidities.
For $\gamma \rightarrow \infty$, where the LL model approaches the
so-called Tonks-Girardeau limit, the bosonic and fermionic models are described by the same set of Bethe equations if the conditions \cite{Iida2005,Wadati2007}:
\begin{equation}\nonumber
N_{F} \leftrightarrow \frac{N_{B}}{2}\,, \quad c_{F} \leftrightarrow - 2c_{B} \,,  \quad \gamma_{F} \leftrightarrow \frac{\gamma_{B}}{4} \,, \quad m_{F} \leftrightarrow \frac{m_{B}}{2}\,,
\end{equation}
are met.

For the LL model in the thermodynamic limit, one can write an integral equation for the density of pseudorapidities, $f_{B}(x)$, given by \cite{Lieb63}:
\begin{equation} \label{Love1}
\frac{f_{B}(x)}{2} - \frac{1}{2\pi}\int_{-B_{B}}^{B_{B}}\frac{f_{B}(y)}{1 + (y-x)^{2}}\,dy = 1\,,
\end{equation}
while the dimensionless coupling constant is given by: 
\begin{equation} \label{gammaB}
\frac{1}{\gamma_{B}} = \frac{1}{4\pi} \int_{-B_{B}}^{B_{B}}f_{B}(x) \,dx \,.
\end{equation}
In \eqref{gammaB} the pseudorapidities take values in $[-B,B]$, $B$ being the largest pseudorapidity.
If one defines the (dimensionless) ground state energy density as: $e(\gamma) = \frac{2m}{\hbar^2} \left(\frac{E}{L}\right) n^{-3}$, where $E$ is the ground state energy of the system, then the energy density function is related to the weight function $f_B(x)$ as \cite{Lieb63}:
\begin{equation} \label{egammaB1}
e_{B}(\gamma_{B}) = \frac{\gamma_{B}^{3}}{4\pi} \int_{-B_{B}}^{B_{B}} x^{2}f_{B}(x)\,dx\,.
\end{equation}  

One can write analogous equations for the GY model. The density of pseudorapidities, here denoted by $f_{F}(x)$, is given by:
\begin{equation}\label{Love2}
\frac{f_{F}(x)}{2} + \frac{1}{2\pi}\int_{-B_{F}}^{B_{F}}\frac{f_{F}(y)}{1 + (y-x)^{2}}\,dy = 1 \,,
\end{equation}
while the dimensionless coupling reads: 
\begin{equation} \label{gammaF}
\frac{1}{\gamma_{F}} = \frac{1}{\pi} \int_{-B_{F}}^{B_{F}}f_{F}(x) \,dx \,. 
\end{equation}
From these two equations, one can express the
dimensionless ground state energy as: 
\begin{equation}  \label{egammaF1}
e_{F}(\gamma_{F}) = -\frac{\gamma_{F}^{2}}{4} + \frac{\gamma_{F}^{3}}{\pi} \int_{-B_{F}}^{B_{F}} x^{2}f_{F}(x)\,dx\,.
\end{equation}
Note that in the weak coupling limit, $\gamma\rightarrow 0$, the integral equations for the weight functions become singular, and solving them for a weakly interacting dilute gas is not a straightforward task both analytically and numerically \cite{Hutson63,Lang-Minguzzi}.

For each model, these three equations — for the density of pseudorapidities,
the pseudorapidity cut-off, and the dimensionless ground state energy — must be solved numerically at any given coupling strength, and no single closed-form, analytical expression for these quantities is available across the whole range
of interactions. Let us then now discuss how to solve them numerically.

\section{Numerical solution of the Love equation}
\label{sec:numsol}
In this Section, we briefly discuss the numerical solution of the Love integral equation and present numerical results that will be used in the following.
First of all, let us define, for convenience, the parameter $\lambda \equiv 1/B$. Eqs. (\ref{gammaB}) and (\ref{gammaF}) will then be expressed as:
\begin{equation}\label{lambda_B}
\lambda_{B}= \frac{\gamma_{B}}{4\pi}\int_{-1}^{1}f_{B}(B_{B}t)\,dt\,,
\end{equation}
and
\begin{equation}\label{lambda_F}
\lambda_{F} = \frac{\gamma_{F}}{\pi}\int_{-1}^{1}f_{F}(B_{F}t)\,dt\,,
\end{equation}
where $t = x/B$ are the normalized pseudorapidities. Then, rewriting the density functions:
\begin{eqnarray}
\nonumber
g_{B}(t) = f_{B}(B_{B}t)\,,\\\nonumber
g_{F}(t) = f_{F}(B_{F}t)\,,
\end{eqnarray}
for the LL model and the GY model, respectively, we can rewrite Eqs. (\ref{Love1}) -- (\ref{egammaB1}), for the LL model, as: 
\begin{gather}
\nonumber 
g_{B}(t) - \frac{\lambda_{B}}{\pi}\int_{-1}^{1}\frac{g_{B}(t')}{\lambda_{B}^{2} + (t' - t)^{2}} \,dt' = 2\,,\\\nonumber
\lambda_{B} = \frac{\gamma_{B}}{4\pi}\int_{-1}^{1}g_{B}(t)\,dt\,,\\\nonumber
e_{B}(\gamma_{B}) = \frac{\gamma_{B}^{3}}{4\pi \lambda_{B}^{3}}\int_{-1}^{1} t^{2}g_{B}(t)\,dt\,,
\end{gather}
while Eqs. (\ref{Love2}) -- (\ref{egammaF1}) of the GY model, as:
\begin{gather}\nonumber
g_{F}(t) + \frac{\lambda_{F}}{\pi}\int_{-1}^{1}\frac{g_{F}(t')}{\lambda_{F}^{2} + (t' - t)^{2}}\, dt' = 2\,,\\\nonumber
\lambda_{F} = \frac{\gamma_{F}}{\pi}\int_{-1}^{1}g_{F}(t)\,dt\,,\\\nonumber
e_{F}(\gamma_{F}) = -\frac{\gamma_{F}^{2}}{4} +  \frac{\gamma_{F}^{3}}{\pi \lambda_{F}^{3}}\int_{-1}^{1} t^{2}g_{F}(t)\,dt\,.
\end{gather}
These equations can be solved in several ways. The most direct method employs Chebyshev quadrature with $M$ points (we take $M=100$). Denoting by $h(t)$ a generic function and by $M$ the number of quadrature points, one may rely
on the following approximation \cite{DelvesBook}:
\begin{equation}
\nonumber
\int_{-1}^{1} h(t)\, dt \, \simeq \, \sum_{j=1}^{M}h(t_{j})  w_{j} \sqrt{1 - t_{j}^{2}}\,,
\end{equation}
where $t_{j} = \cos\left( \frac{(2j - 1)\pi}{2M} \right)$, and the weights for the Chebyshev quadrature reads $w_{j} \equiv \frac{\pi}{M}$, $ \forall j$. The first two equations of each model must be then solved iteratively. 

Another method was proposed by Love, in his original work \cite{Love49}. He showed that, for fixed $\lambda$, the weight function can be written as:
\begin{equation}
g_{B/F}(t) = 1 + \sum_{n=1}^{\infty} (\pm 1)^{n}\int_{-1}^{1}K_{n}(t',t)\, dt'\,,
\end{equation}
where:
\begin{eqnarray}
\nonumber
K_{1}(t',t) = \frac{1}{\pi}\frac{\lambda}{\lambda^{2} + (t'-t)^{2}}\,,\\\nonumber
K_{n}(t',t) = \int_{-1}^{1} K_{n-1}(t,s)K_{1}(s,t') ds\,,
\end{eqnarray}
where, again, we used the Chebyshev quadrature to compute the integrals. We truncated the series when the terms had an absolute value below $10^{-40}$. Agreement with the ``direct" method is excellent, with discrepancies at most of order $10^{-10}$.

\begin{figure}[h!!]
\includegraphics[width=0.7\columnwidth]{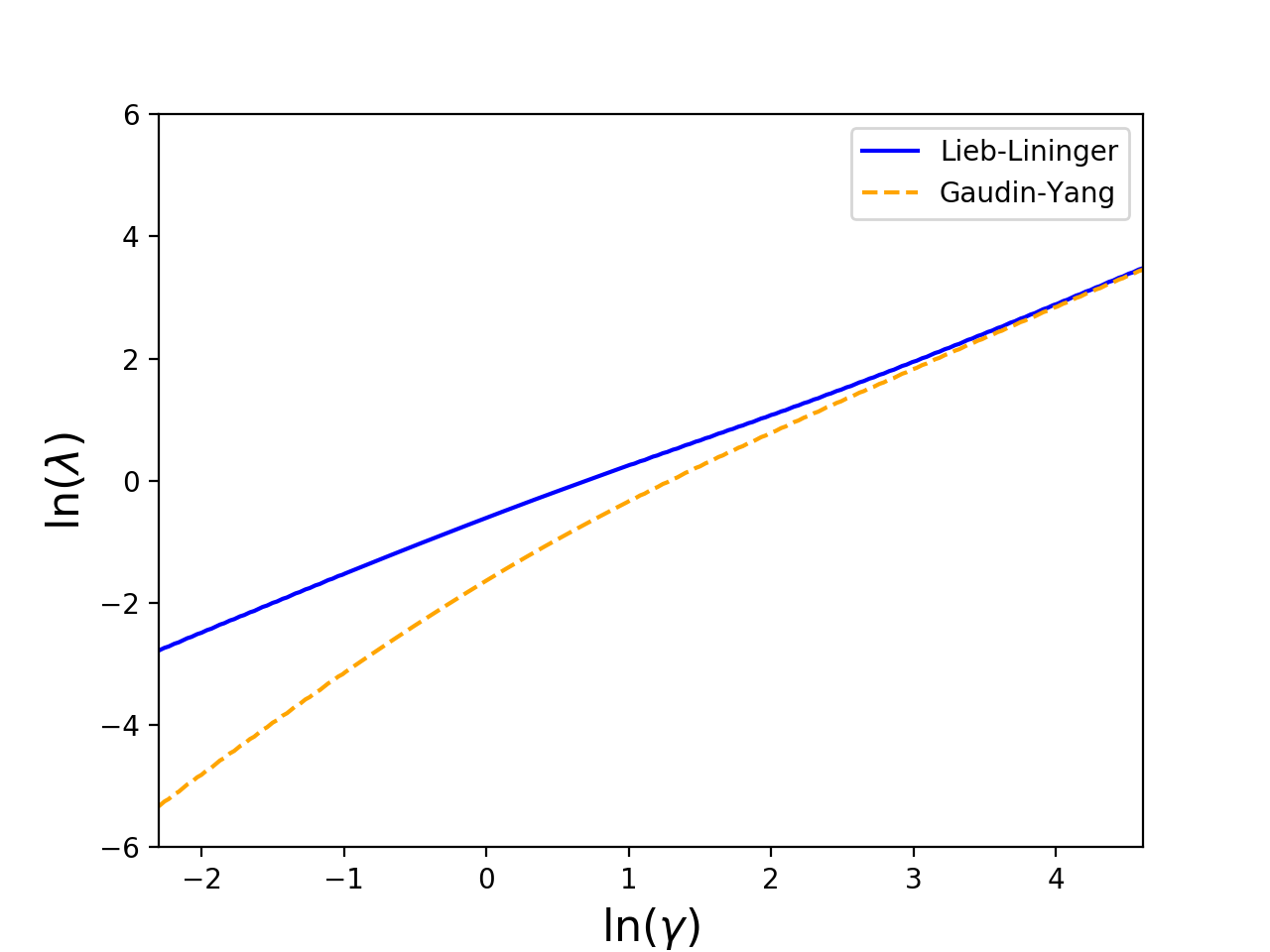}
\caption{Logarithm of the inverse of the cut-off of the pseudorapidities, $\lambda=1/B$, {\it vs} the logarithm of the dimensionless coupling, $\gamma$, for the GY (orange line) and LL (blue line) models.}
\label{fig1}
\end{figure}

\begin{figure}[h!!]
\includegraphics[width=0.7\columnwidth]{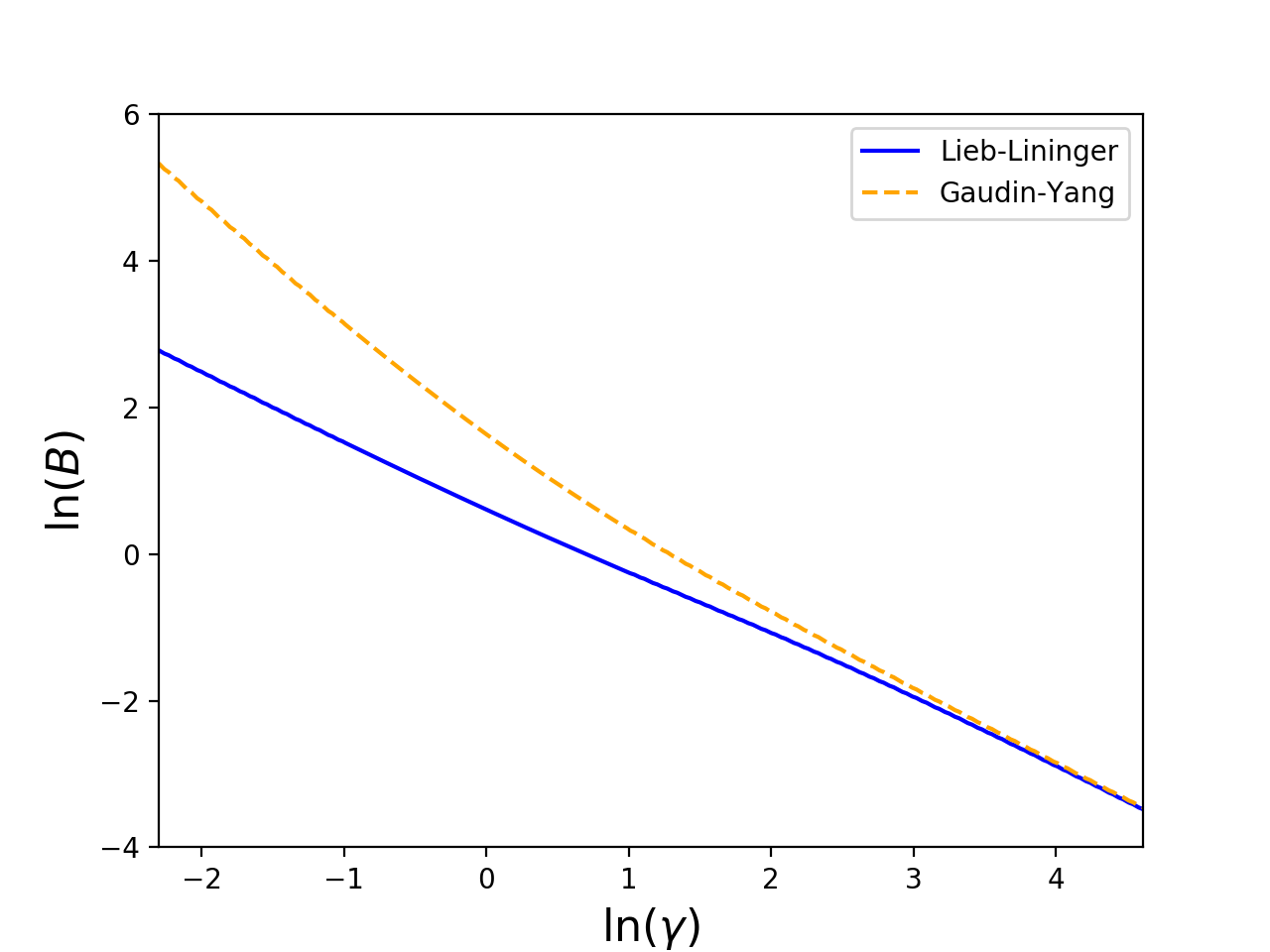}
\caption{Logarithm of the cut-off of the pseudorapidities, $B$, {\it vs} the logarithm of the dimensionless coupling, $\gamma$, for the GY (orange line) and LL (blue line) models.}
\label{fig2}
\end{figure}

These numerical solutions allow us to study the $\gamma$ dependence of the normalized parameters $\lambda$. In Fig. \ref{fig1} we plot the behaviour for $\ln (\lambda_F)$ in orange line, while $\ln (\lambda_B)$ is reported in blue solid line. Note that the two cut-off pseudorapidities coincide at very strong coupling, \textit{i.e.} $\gamma\rightarrow \infty$, where hard--core interactions suppress the effect of statistics properties of the gases and the two models become equivalent. In Fig. \ref{fig2} we plot the behaviour of the cut-off pseudorapidities $B_B$ and $B_F$ against the dimensionless coupling strength.

It is convenient to define a normalized density of pseudorapidities, denoted by $\tilde{f}(t)$:
\begin{equation}
\tilde{f}_{B}(t) = \frac{1}{4\pi}\frac{\gamma_{B}}{\lambda_{B}}f_{B}(B_{B}t)\,,
\end{equation}
for the LL model, whereas for the GY model:
\begin{equation}
\tilde{f}_{F}(t) = \frac{1}{\pi}\frac{\gamma_{F}}{\lambda_{F}}f_{F}(B_{F}t)\,.
\end{equation}

With these notations, Eqs. (\ref{gammaB}) and (\ref{gammaF}) become the condition of normalization, valid in both cases:
\begin{equation} \label{norm_eq}
\int_{-1}^{1} \tilde{f}(t)\,dt  = 1\,.
\end{equation}
On the other hand, for the LL model Eqs. (\ref{Love1}) and (\ref{egammaB1}) become, respectively:
\begin{equation} \label{fBtilde}
\tilde{f}_{B}(t) - \frac{\lambda_{B}}{\pi}\int_{-1}^{1}\frac{\tilde{f}_{B}(t')}{\lambda_{B}^{2} + (t' - t)^{2}} \,dt' = \frac{1}{2\pi}\frac{\gamma_B}{\lambda_{B}}\,,
\end{equation}  
and:
\begin{equation} \label{egammaB2}
e_{B}(\gamma_B) = \left(\frac{\gamma_B}{\lambda_{B}}\right)^2\,\int_{-1}^{1} t^{2}\tilde{f}_{B}(t)\,dt\,.
\end{equation}  

For the GY model we get that Eqs. (\ref{Love2}) and (\ref{egammaF1})
transform into:
\begin{equation} \label{fFtilde}
\tilde{f}_{F}(t) + \frac{\lambda_{F}}{\pi}\int_{-1}^{1}\frac{\tilde{f}_{F}(t')}{\lambda_{F}^{2} + (t' - t)^{2}} \,dt' = \frac{2}{\pi}\frac{\gamma_F}{\lambda_{F}}\,,
\end{equation}  
and the dimensionless ground state energy density is given by:
\begin{equation} \label{egammaF2}
e_{F}(\gamma_F) = -\frac{\gamma_F^{2}}{4} + \left(\frac{\gamma_F}{\lambda_{F}}\right)^2\,\int_{-1}^{1} t^{2}\tilde{f}_{F}(t)\,dt\,.
\end{equation}  

It is worth emphasizing once more the close similarity between
Eqs. (\ref{fBtilde}) and (\ref{fFtilde}) and the Love equations for the condenser problem in, respectively, (\ref{two_plates_LL}) and (\ref{two_plates_GY}),
once $\mathcal{Z}=1$ has been chosen and normalized by $2V_1 /\pi$.
They are integral equations of the same type, but where the distance between the plates and the capacitance are replaced by the coupling strength parameter and the cut-off of the pseudorapidities of the systems. As noted above, these equations cannot be solved analytically over the whole range of interactions, and one must resort to numerical approaches, which can be time-consuming, particularly when $e(\gamma)$ is required at many different values of the coupling. With the aim of obtaining an expression for the ground state energy density valid in all regimes of interactions, we will now discuss how to build an interpolation scheme in order to write the density of pseudorapidities in terms of Chebyshev polynomials. This will be the basis from which one can then construct a Hermite interpolation also for the function $e(\gamma)$, which has to be compared with the numerical solution which will be used for quality checks.

\section{Chebyshev interpolation for the solution of Love equations}
\label{sec:5}
In order to find analytical approximate expressions for the normalized solution to the Love equations (\ref{fBtilde}) and (\ref{fFtilde}), we propose to use a Hermite interpolation via Chebyshev polynomials. The Chebyshev polynomials in the closed interval $x \in [-1,1]$, are defined as: 
\begin{equation}\nonumber
T_n(x) = \cos \left[n \cos^{-1}(x)\right]\,,\quad \quad n \in \mathbb{N} \,.
\end{equation}

The extrema of the Chebyshev polynomials, commonly referred to as \textit{Chebyshev points}, play an important role in approximation theory, where they are used as interpolation nodes to minimize the interpolation error \cite{LevyBook}. The first $M$ Chebyshev points for the polynomial $T_{2M}(x)$ are located at positions: 
\begin{equation}\label{chebypoints}
t_{i} = \cos\left(i\,\frac{\pi}{2M}\right)\,, \quad \quad i = 0,1,2,\dots,M-1\,.
\end{equation}

By Hermite interpolation we mean that, in addition to matching the values of the function at the Chebyshev points, we also match its derivatives at, in general, a number of further points. In our specific case, for a certain fixed $\gamma$, the interpolation procedure will be the same for the solutions of both types of Love equations.

Let us present our method. First of all, we notice that $\tilde{f}(t)$ is an even function in $t$, hence we can interpolate it with a polynomial with even powers of $t$. This property also allows us to consider only the interval $t \in [0,1]$. To find a good approximation, we selected the first $M$ Chebyshev points of the Chebyshev polynomial of degree $2M$ as in Eq. (\ref{chebypoints}).
Then, in order to fix the values of the function at the interpolating Chebyshev points, we used the numerical results of $\tilde{f}$ obtained with the methods seen in the previous Section for each $t_{i}$. In order to implement the Hermite interpolation, we evaluated the derivative of the function at two Chebyshev points far from $t=1$ by using the central finite difference. The two points at which the derivatives are imposed, which we call Hermite points, were chosen so as to minimize the error in $\tilde{f}'(t)$ over the range $t\in[0,1]$. Moreover, in order to check for convergence in the value of the numerical derivative obtained via finite differences, we used the central finite difference formula up to the $16$-th order of accuracy. Finally, we approximated the solutions of Love equations as: 
\begin{equation}\label{interp_ftilde}
\tilde{f}(t) = \sum_{i=0}^{M+1} \tilde{C}_{i}t^{2i}\,,
\end{equation} 
where $M$ coefficients $\tilde{C}_{i}$ are fixed by using the previously obtained values of $\tilde{f}(t_i)$, while the other $2$ coefficients are fixed thanks to the Hermite points for the derivatives of the functions.

\begin{figure*}[h!!]
\centering
\includegraphics[width=0.45\columnwidth]{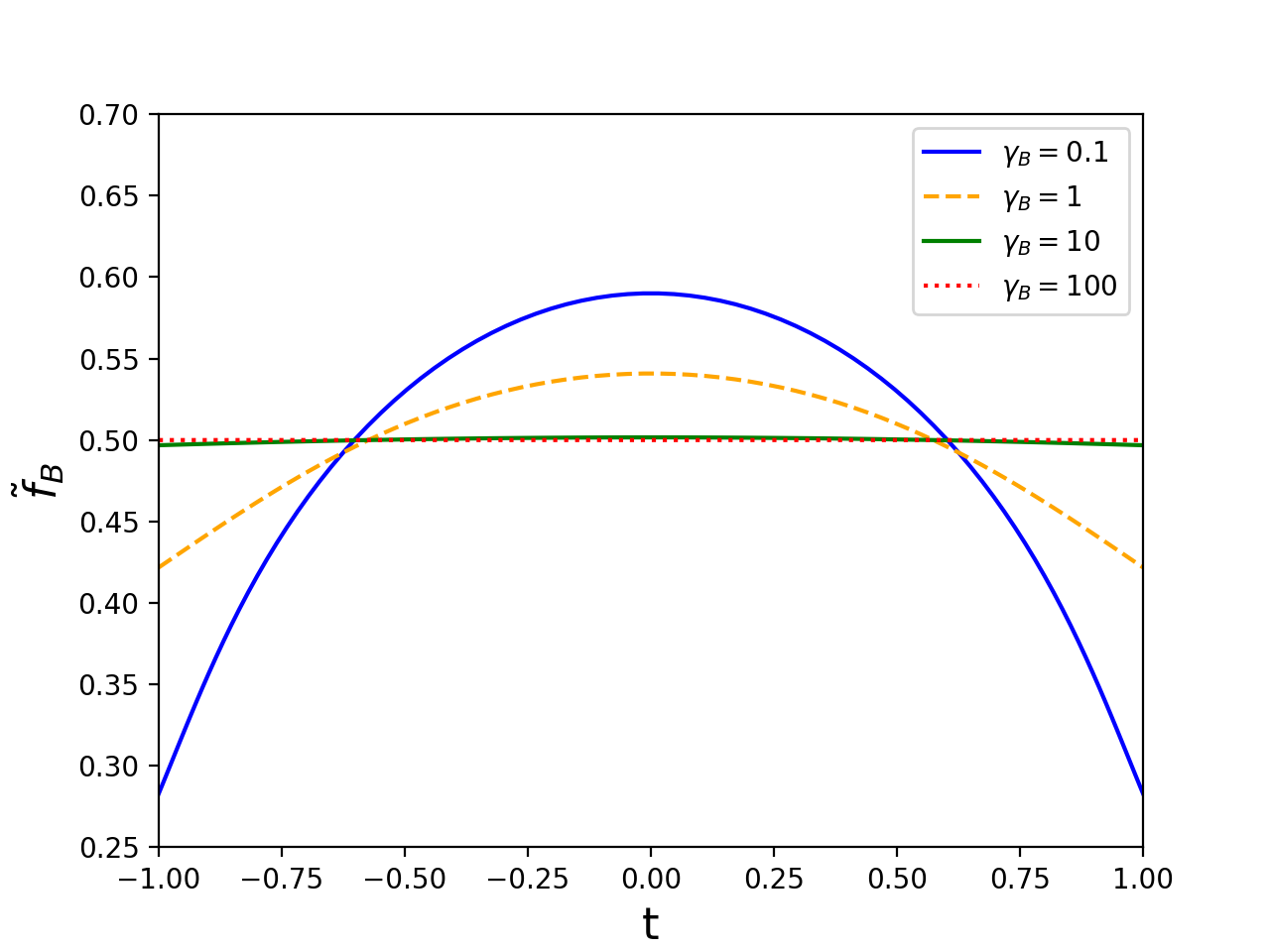}
\includegraphics[width=0.45\columnwidth]{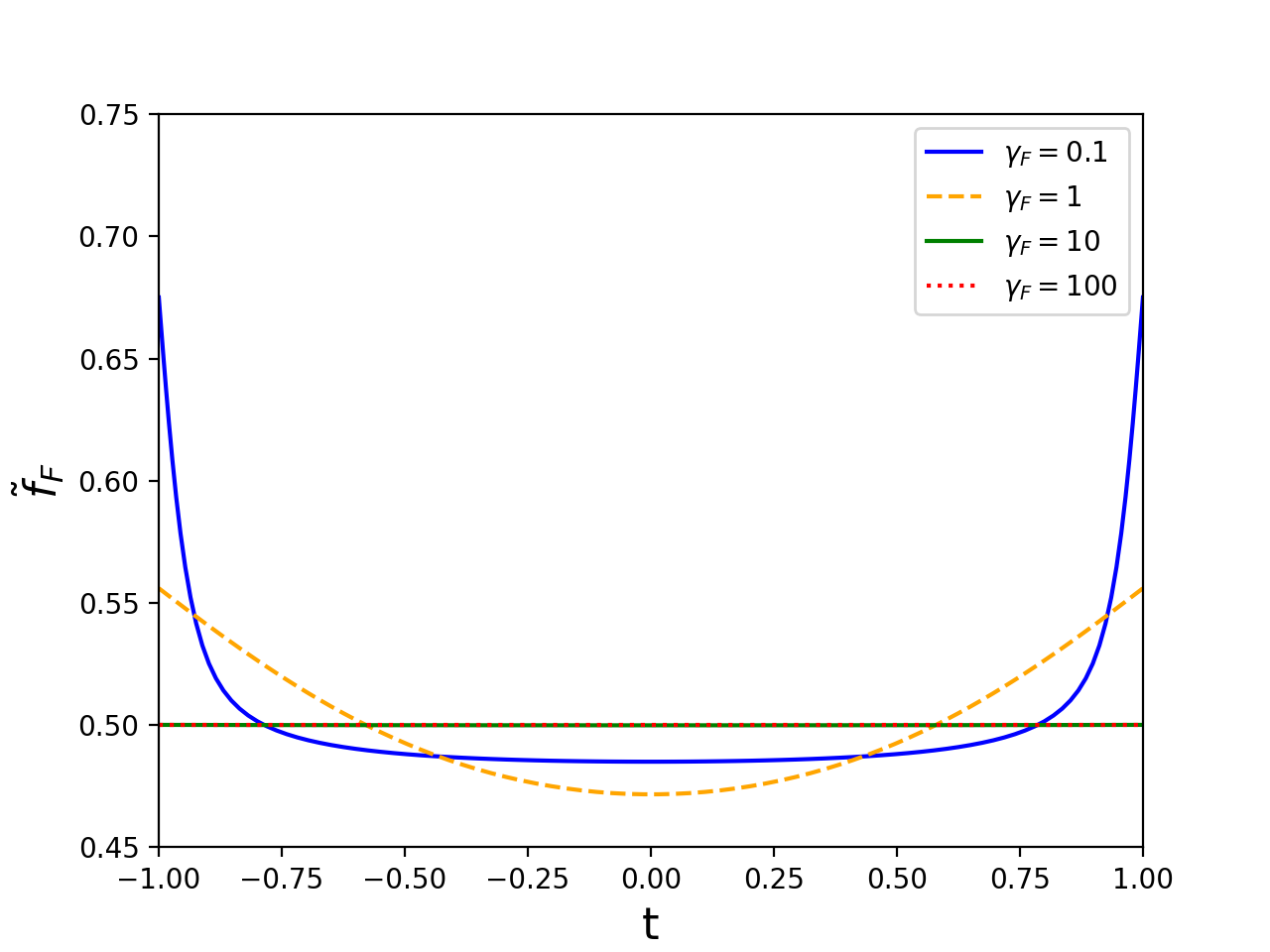}
\caption{The weight function $\tilde{f}(t)$, normalized solution to the Love equation of the first kind (left) and second kind (right), given by the Chebyshev interpolation, for different values of the interaction parameter $\gamma$. These figures are the results from using only $17(+2)$ points for the interpolation.}
\label{fig3}
\end{figure*}

\begin{figure*}[h!!]
\centering
\includegraphics[width=0.45\columnwidth]{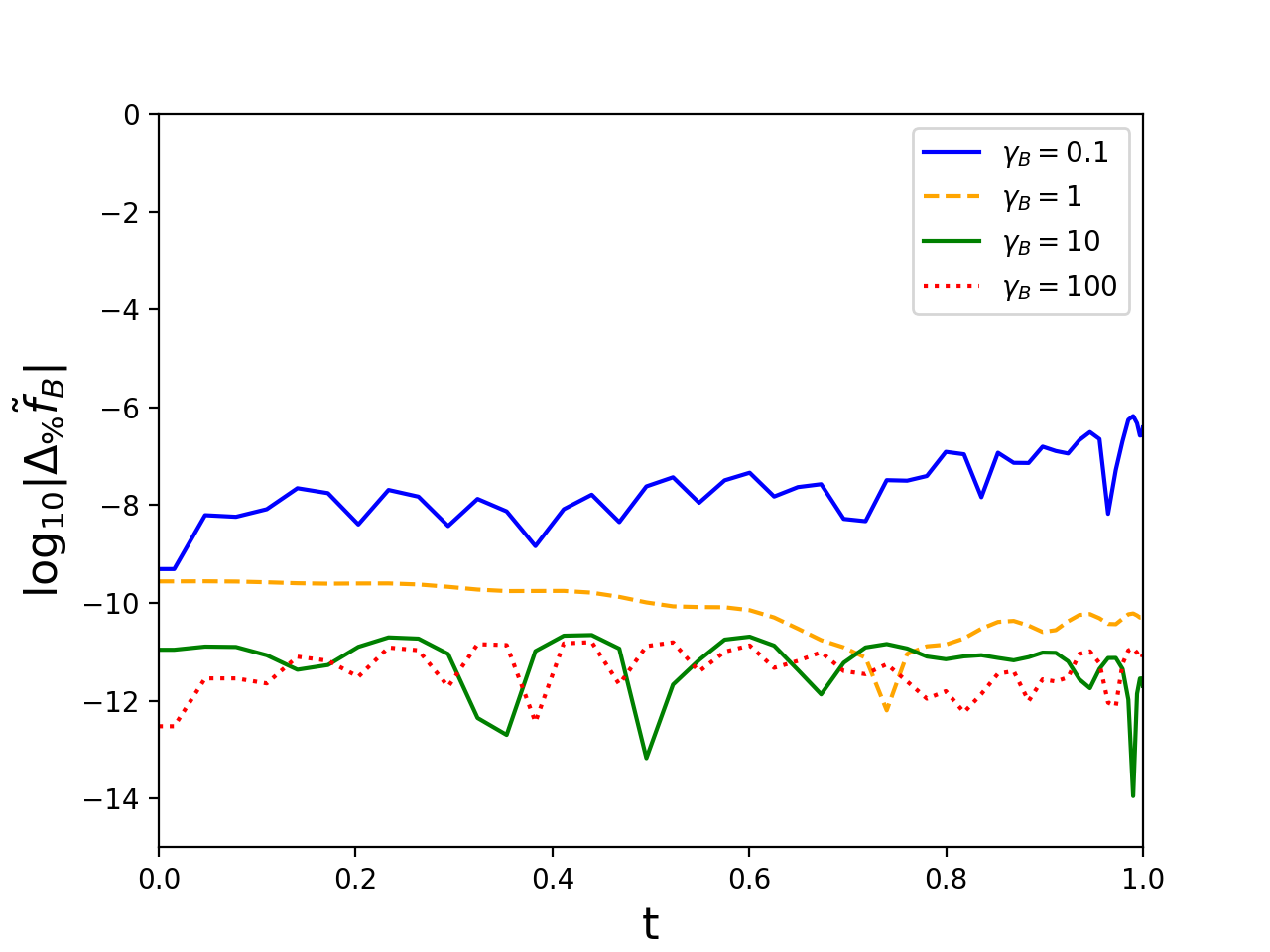}
\includegraphics[width=0.45\columnwidth]{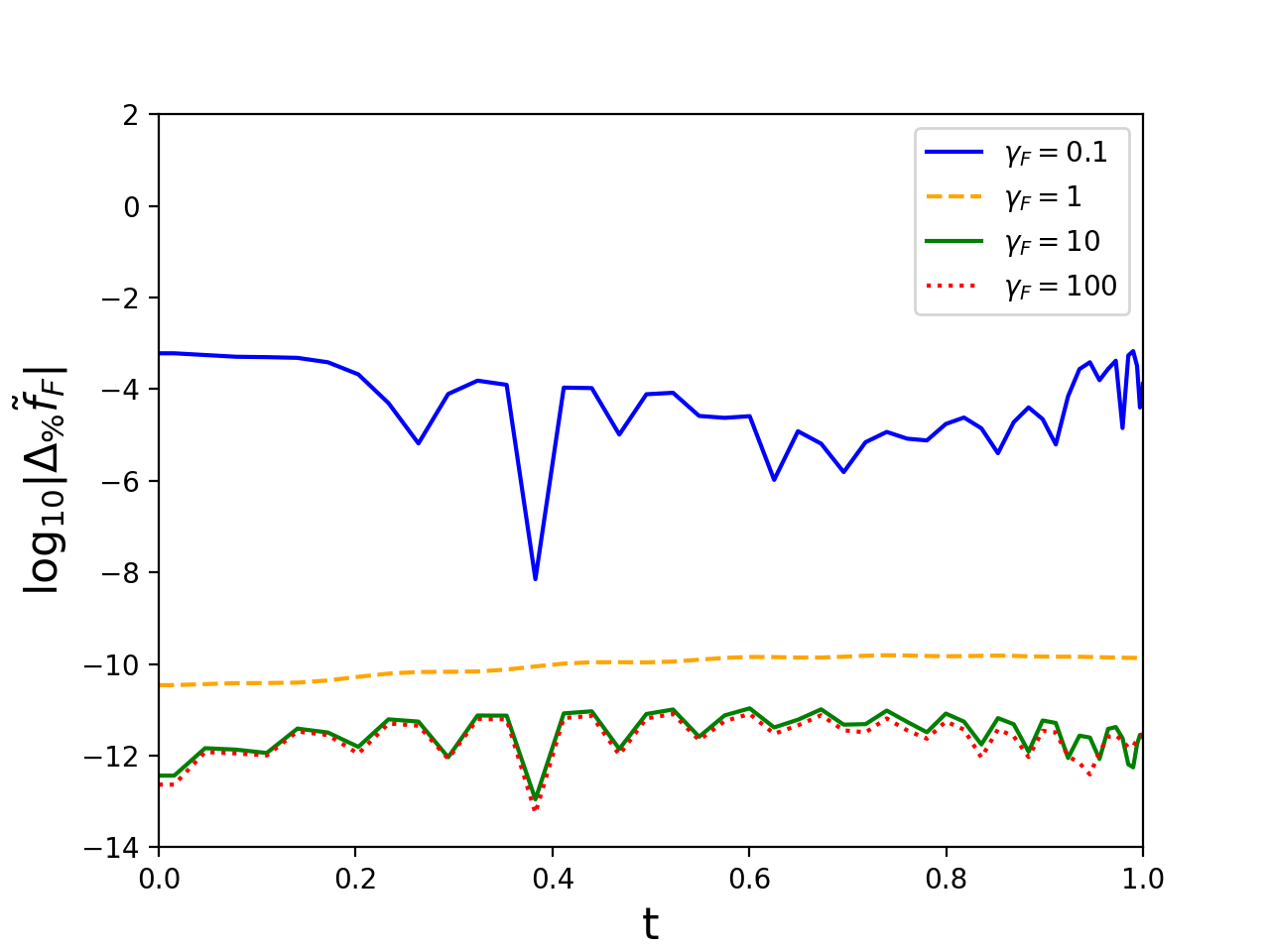}
\caption{Logarithmic behaviour of the absolute value of the relative percentage error of the solutions of Love equation of the first kind (left) and the second kind (right) with respect to $t$, plotted at different values of the interaction parameter $\gamma$. The errors are evaluated from an interpolating scheme which uses only $17(+2)$ points.}
\label{fig4}
\end{figure*}

We have chosen $M = 17$, therefore with $19$ constraints, and the interpolation is then done with a polynomial of maximum degree of $36$. More interpolation points can of course be used, but we have verified that $M=17$ already yields very accurate results, as discussed below. In Fig. \ref{fig3} we plotted the solutions of the Love equations of the first and second kind on the left and right side plot respectively, for different values of the parameter $\gamma$, obtained by the Hermite interpolation scheme described above. They are solutions of Eqs. (\ref{fBtilde}) and (\ref{fFtilde}), having Eq. (\ref{norm_eq}) satisfied. Notice that in the strong coupling limit the two weight functions coincide and reduce to the constant $1/2$, whereas in the weak coupling limit the LL density of pseudorapidities approaches the well-known semicircle law \cite{Lieb63}. In Fig. \ref{fig4} we plot the logarithm of the absolute value of the relative percentage errors of the interpolation scheme with respect to the numerical estimates, denoted by $\log_{10} \left| \Delta_\% \tilde{f}(t)\right|$. 

\section{Chebyshev parametrization for the momenta of Love equations}
\label{sec:6}
The momenta of the solutions of Love equations are also of considerable importance. We have seen for the LL and GY models, for example, that the second momenta of the weight functions of the models are related to the dimensionless ground state energy density, which in turn will determine other important properties of the system, as we are going to discuss in the next Sections. These momenta depend on the coupling constant $\gamma$. In order to obtain accurate analytical approximations, we apply the same Hermite--Chebyshev interpolation to the second, fourth and sixth momenta of Love equations of both kinds, in a certain interval of the coupling constant.

Let us denote the momenta by $\sigma_{i}^{B}$, for Love equation of the first kind, and $\sigma_{i}^{F}$, for Love equation of the second kind, with $i = 2$, $4$ and $6$, \textit{i.e.} in general:
\begin{equation}\label{sigma_iformula}
\sigma_i = \int_{-1}^1 t^i\,\tilde{f}(t)\,dt\,.
\end{equation} 
In order to proceed with the Hermite interpolation via Chebyshev polynomials, we need to introduce an auxiliary variable $\xi \in [-1, 1]$, related to $\gamma$ as:
\begin{equation}\label{csi}
\gamma = 2 \,\frac{A + \xi}{B - \xi}\,,
\end{equation} 
where $A$ and $B$ are constants determined by the range of $\gamma$ to be mapped onto $[-1,1]$. 

As a first illustration, consider an interpolation scheme over the range $\gamma \in [0.1,100]$. Note that reaching lower values, $\gamma<0.1$, is numerically very demanding. With the methods at hand, indeed, convergence is not attained at very small couplings, where instead we will rely on analytical exact expressions, as we are going to see in a moment. 

In order to map the $\gamma$ interval $[0.1,100]$ into the $\xi$ interval $[-1,1]$, one has to choose: $A = \frac{367}{333}$ and $B = \frac{347}{333}$, which can be found from Eq. (\ref{csi}). In this case, since the momenta $\sigma_{i}(\xi)$ are not even functions of $\xi$, the interpolation includes both odd and even powers and we will consider the set of points given by the first $M$ extrema of the Chebyshev polynomials $T_{M}(\xi)$:
\begin{equation}\label{Chebyshev-points}
\xi_{i} = \cos\left(i\,\frac{\pi}{M}\right)\,, \quad \quad i = 0,1,2,\dots,M-1\,,
\end{equation}
in order to set the interpolation points.
Then the Hermite interpolation is done by finding the coefficients $\tilde{C}_{k}^j$ of
\begin{equation}
\sigma_j (\xi) = \sum_{k=0}^{M+1} \tilde{C}_{k}^j \xi^{k}\,,
\end{equation} 
once we fix numerically the values of the momenta at the Chebyshev points, $\sigma_j(\xi_i)$, and once also the first derivatives of the momenta at the end points of the interval, $\sigma'_j(\xi=\pm 1)$, are fixed. We stress that the values of $\sigma_j$ at the Chebyshev points are obtained by numerically evaluating Eq. (\ref{sigma_iformula}) using Chebyshev quadrature method and where the values of the density of pseudorapidities $\tilde{f}$ are obtained with the methods presented in Section \ref{sec:numsol}. Finally, the values for the derivatives $\sigma'_j$, are obtained numerically via backward and forward finite differences.

\begin{figure*}[h!!]
\centering
\includegraphics[width=0.45\columnwidth]{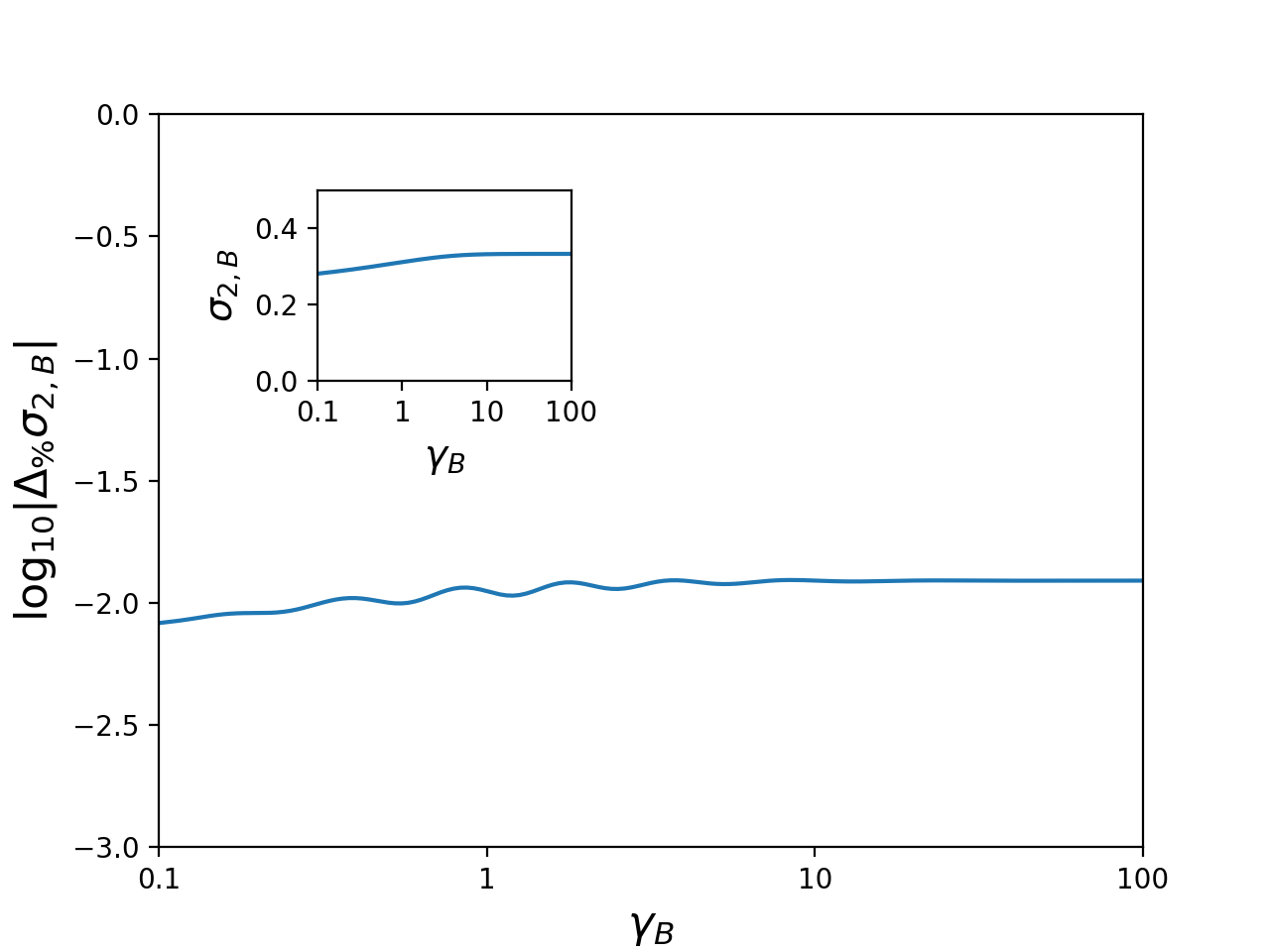}
\includegraphics[width=0.45\columnwidth]{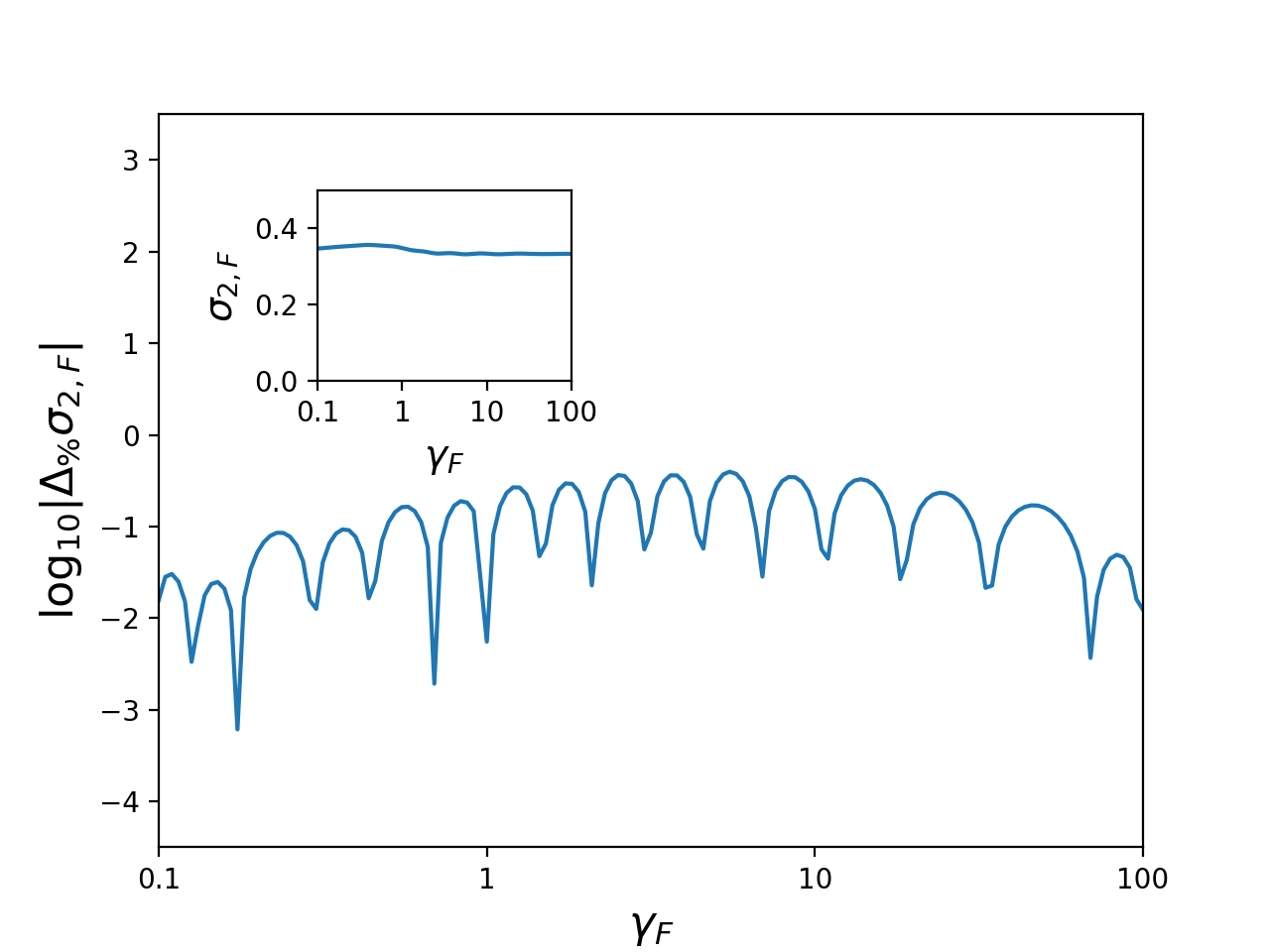}
\caption{Logarithm of the absolute value of the percentage error for the second momentum function of Love equation of the first (left) and second kind (right) with respect to the coupling strength, $\gamma$, which is reported in log scale. The errors are obtained by comparing the results obtained via Chebyshev interpolation with the numerical outcomes. The insets show $\sigma_{2}$ {\it vs} $\gamma$ in linear-log scale. These results are obtained from the Hermite--Chebyshev interpolation scheme with $M=17$.}
\label{fig5}
\end{figure*}

\begin{figure*}[h!!]
\centering
\includegraphics[width=0.45\columnwidth]{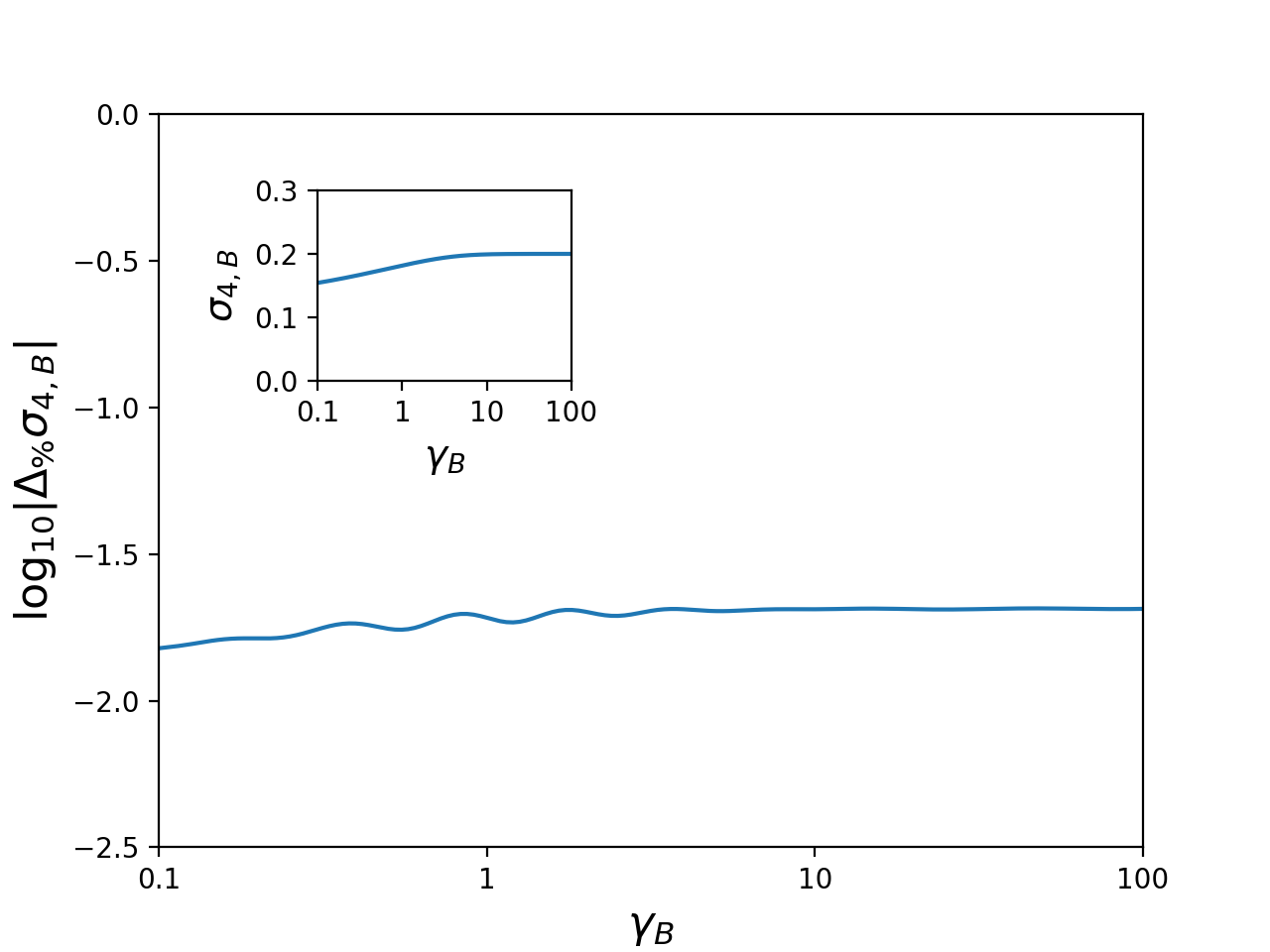}
\includegraphics[width=0.45\columnwidth]{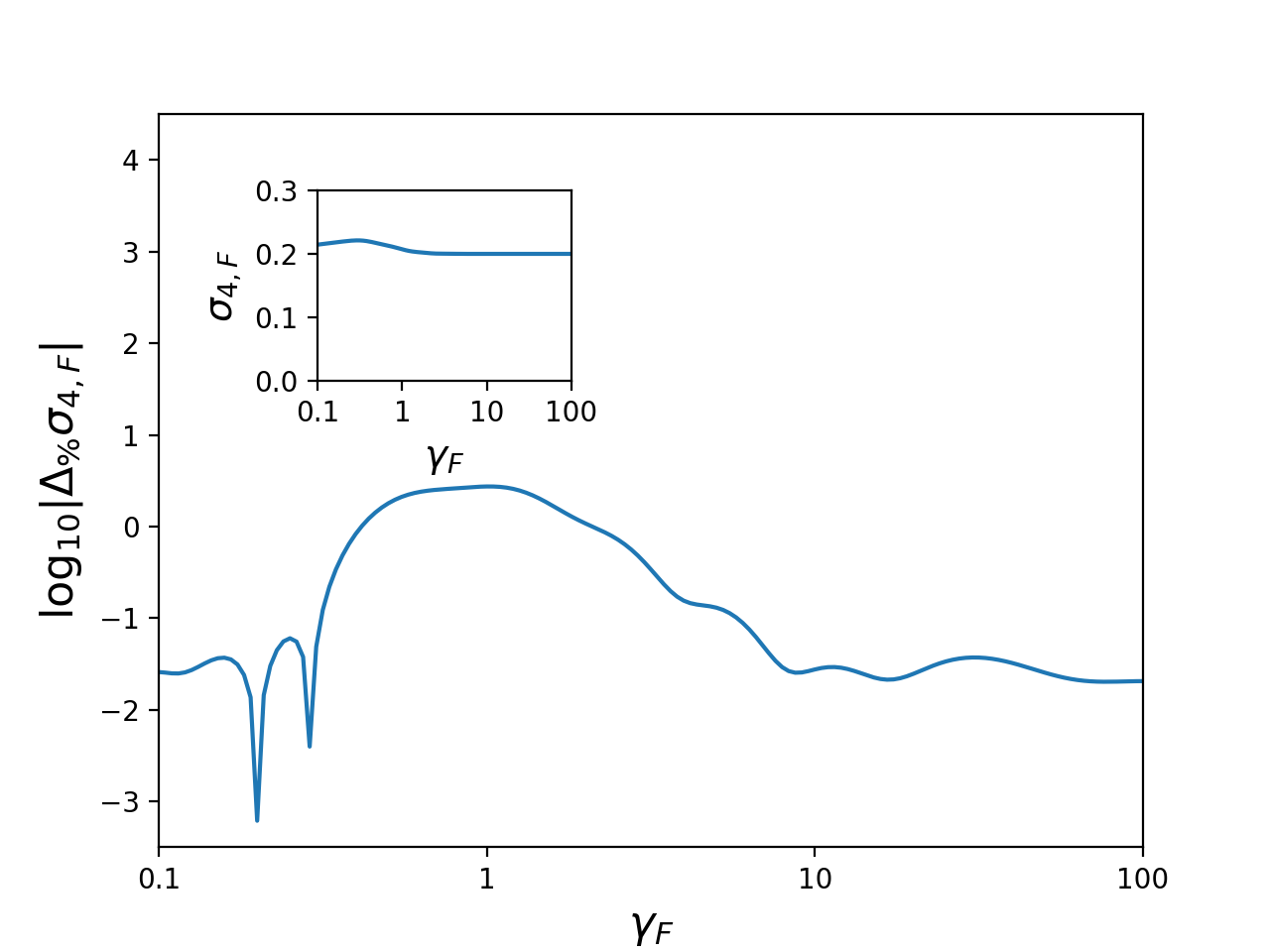}
\caption{Logarithm of the absolute value of the percentage error for the fourth momentum function of Love equation of the first (left) and second kind (right) with respect to the coupling strength, $\gamma$, again reported in log scale. The errors are evaluated by comparing the results obtained via Chebyshev interpolation with the numerical outcomes. In the insets we report the behaviour of $\sigma_{4}$ {\it vs} $\gamma$ in linear- log scale. These results are obtained from the Hermite--Chebyshev interpolation scheme with $M=17$.}
\label{fig6}
\end{figure*}

\begin{figure*}[h!!]
\centering
\includegraphics[width=0.45\columnwidth]{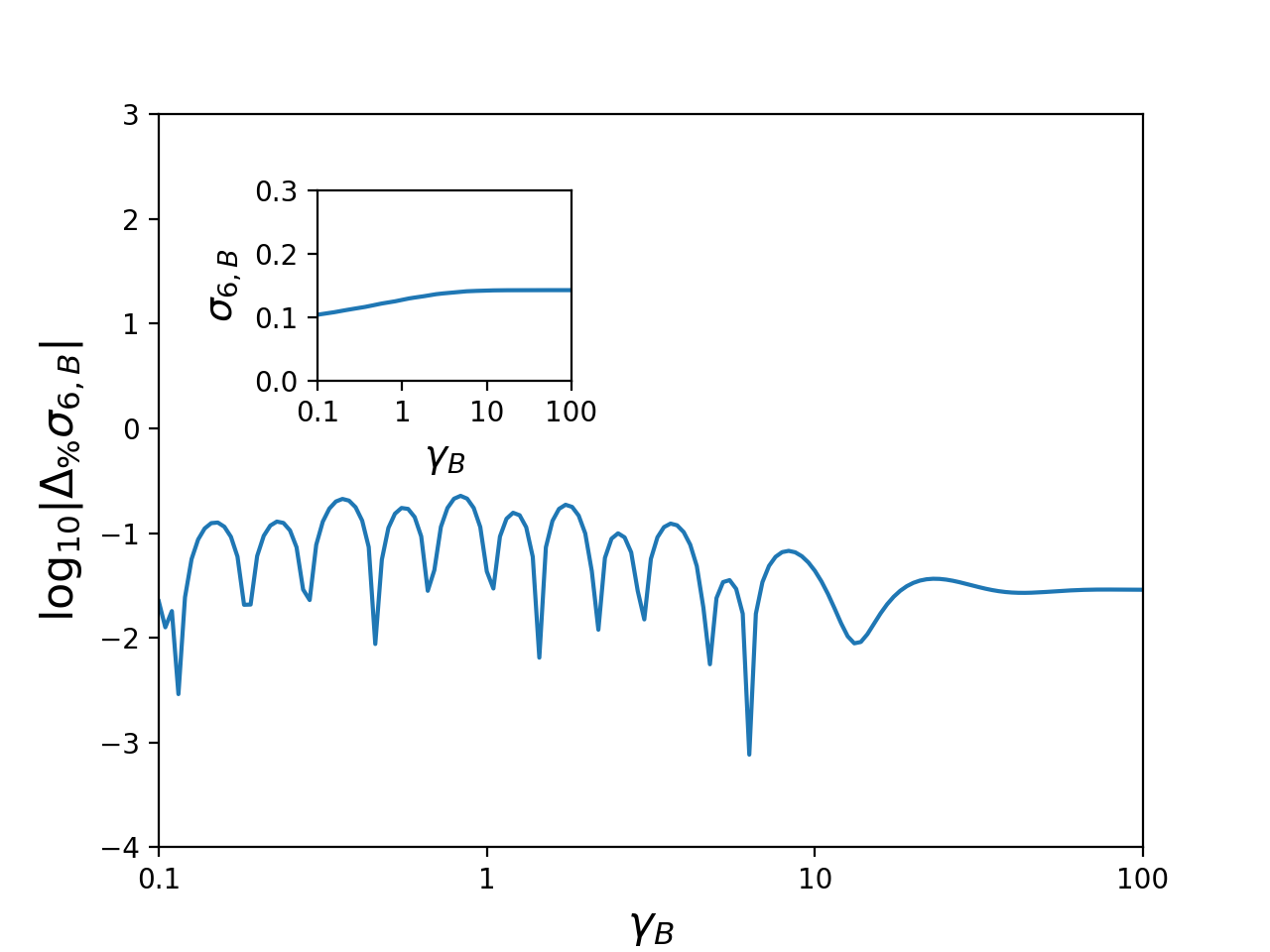}
\includegraphics[width=0.45\columnwidth]{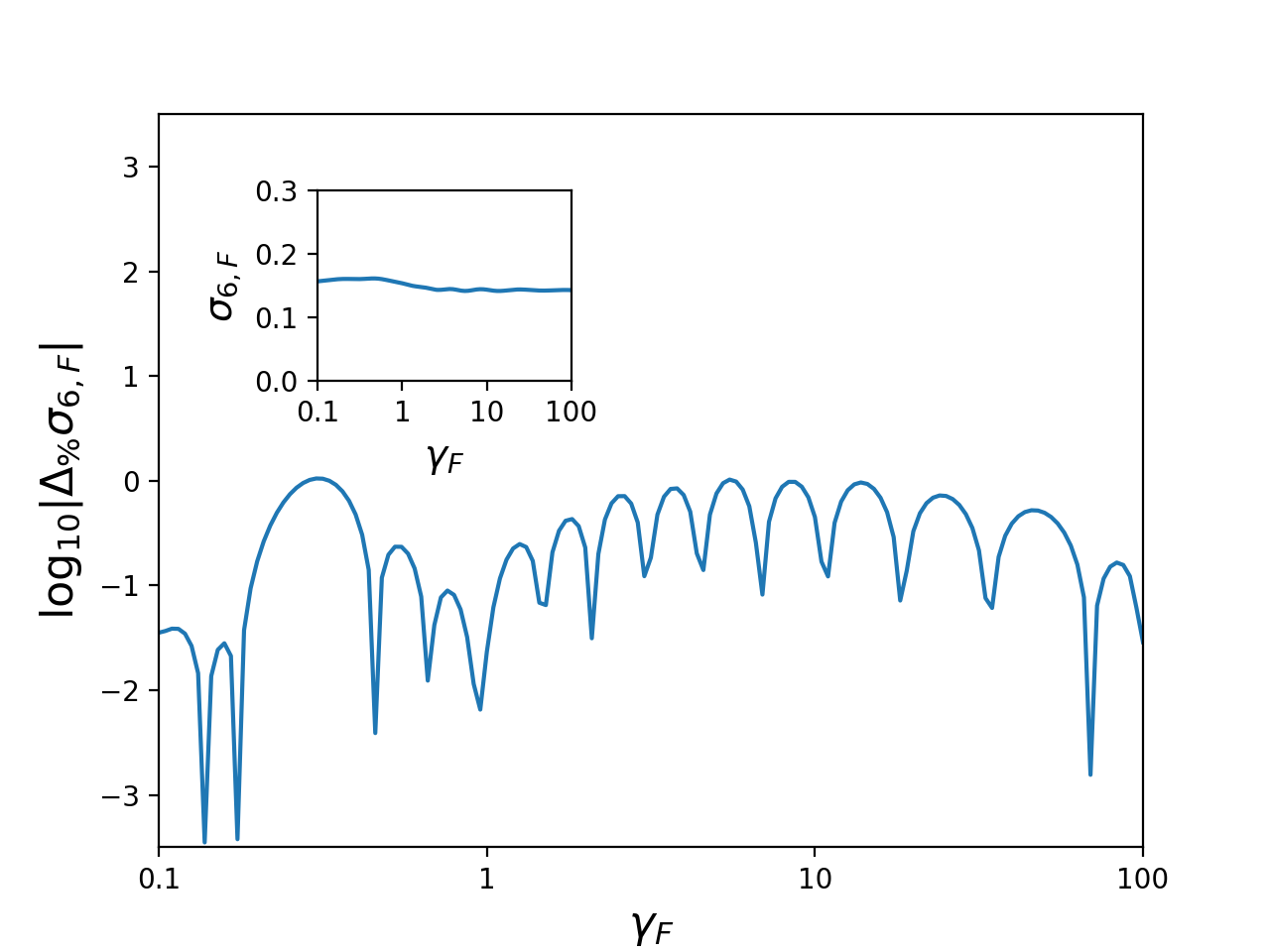}
\caption{Logarithm of the absolute value of the percentage error for the sixth momentum function of Love equation of the first (left) and second kind (right) with respect to the coupling strength, $\gamma$, in log scale. The errors are computed by comparing the results obtained via Chebyshev interpolation with the numerical outcomes. The insets show $\sigma_{6}$ {\it vs} $\gamma$ in linear- log scale. These results are obtained from the Hermite--Chebyshev interpolation scheme with $M=17$.}
\label{fig7}
\end{figure*}

Using only $17(+2)$ points, \textit{i.e.} $M=17$, we obtained the results reported in Fig. \ref{fig5} for the second momentum function of both Love integral equations and the behaviour of the interpolation relative percentage error estimated with respect to the numerical solution. In Figs. \ref{fig6} and \ref{fig7} we plotted the behaviours of the fourth and sixth momenta for both kinds of Love equations, together with their relative percentage errors. In summary, this interpolation scheme yields expressions for the second momenta accurate to about $10^{-2}\%$, whereas for the fourth and sixth momenta the error accumulation is more pronounced, and one finds that the maximum relative percentage error of the Chebyshev parametrization is about $3\%$.

This illustrative example shows how to treat the case $\gamma \in [0.1,100]$; one may then ask whether the scheme can be extended to a wider range of couplings. Besides the failure of the numerical algorithm to converge at very weak coupling, the resulting momentum functions are not sufficiently accurate. The relative percentage errors will run between $10^{-2}\%$ and $10\%$, making this rather simple interpolation procedure not really useful. 

This motivates a central idea of the present work: To divide the range of coupling strengths into three domains, relying in each on either numerical or exact analytical results, where available. It is indeed well known in the literature \cite{Wadati2002,Iida2005,Wadati2007,Lang-Minguzzi,Lang2019,Marino2019} that there exist expressions valid in the weak and strong coupling regimes for the weight functions and the second momenta of Love equations of both kinds. We will focus on the dimensionless energy density function of the LL model, which is related to the second momenta of Love equation of the first kind, as we have seen previously. The same approach could be extended to Love equation of the second kind, \textit{i.e.} to the energy density function of the GY model. Our idea is then to rely on analytical expressions in the regime of weak couplings (where the numerical approach fails) and strong couplings, while we build the above mentioned Hermite--Chebyshev interpolation scheme for $e_B(\gamma_B)$ in the intermediate region of interaction strengths. In order to get a smooth and complete form of the energy density function, we will need to first determine the optimal widths of the three coupling regions, and then match the behaviour of $e_B(\gamma_B)$ and its first derivatives at the matching points of the different coupling regions. In the following Section we will present this method in more detail and we will show the remarkably low values of the relative percentage errors for the analytical approximation of the dimensionless energy density function.

\section{Chebyshev parametrization for the dimensionless ground state energy for the Lieb--Liniger model}
\label{sec:7}
Focusing on Love equation of the first kind, \textit{i.e.} the LL model, exact expressions for the dimensionless energy density valid at high or low couplings are available in the literature. In the regime of weak interactions, \textit{i.e.} for small values of $\gamma_B$, it can be written as \cite{Marino2019,Ristivojevic2019}:
\begin{eqnarray} \nonumber
e_{B}^{w}(\gamma_{B}) &=& \gamma_{B} - \frac{4 \,\gamma_{B}^{3/2}}{3\pi} + \left( \frac{1}{6} - \frac{1}{\pi^{2}} \right)\gamma_{B}^{2} + \frac{3\zeta(3) - 4}{8\pi^{3}}\,\gamma_{B}^{5/2} + \frac{3\zeta(3) - 4}{24\pi^{4}}\,\gamma_{B}^{3} + \frac{60\zeta(3) - 45\zeta(5) - 32}{1024\pi^{5}}\,\gamma_{B}^{7/2}  + \\\nonumber
&&+ \frac{3[4\zeta(3) + 6\zeta(3)^{2} - 15\zeta(5)]}{2048\pi^{6}}\,\gamma_{B}^{4}  + \frac{-4368\zeta(3) + 6048\zeta(3)^{2} + 2520\zeta(5) - 8505\zeta(7) + 1024}{786432\pi^{7}}\,\gamma_{B}^{9/2}+O(\gamma_B^5)\,,\\
\label{low_gamma_Marino}
\end{eqnarray}
where $\zeta(x)$ is the Riemann zeta function, and the superscript $w$ in $e_B^w(\gamma_B)$ stands for ``$weak$". By comparing the above analytical expression with the numerical results, we were able to fix the limit of the first region of the coupling strength (the weak coupling region) at: 
\begin{equation}
\gamma_{B}^w = 0.33\,.
\end{equation}
This value was chosen because the absolute percentage error between the numerical estimate and the one obtained by the aforementioned power-series expansion in $\gamma_B$, reaches a value of: $\Delta_{\%}(e_B(\gamma_{B}^w)) \simeq 1.55 \cdot 10^{-6}$, which is the minimum error value for low coupling strengths.

At strong coupling, \textit{i.e.} large values of $\gamma_B$, the energy density is described by a power-series in $1/\gamma_{B}$ \cite{Lang-Minguzzi}: 
\begin{equation} \label{high_gamma_e}
e_{B}^{s}(\gamma_{B}) = \frac{\pi^{2}}{3} - \frac{4\pi^{2}}{3}\frac{1}{\gamma_{B}} + 4\pi^{2}\frac{1}{\gamma_{B}^{2}} + \left( \frac{32\pi^{4}}{45} - \frac{32\pi^{2}}{3}\right)\frac{1}{\gamma_{B}^{3}}+O(\gamma_B^{-4})\,,
\end{equation}
where the superscript $s$ in $e_B^s(\gamma_B)$ stands for ``$strong$".

We next determine the value of the upper limit of $\gamma_B$ above which the analytical expression (\ref{high_gamma_e}) is used in the three-range interpolation scheme. We fixed an upper value of $\gamma_{B}^s = 1010$ by noting that an optimal, \textit{i.e.} minimum, value of the error $\Delta_{\%}(e_{B}(\gamma_{B}^s)) \simeq 1.61\cdot10^{-6}$ evaluated between the numerical computation and the power-series result (\ref{high_gamma_e}) is reached.

To further check the reliability of the coupling strength intervals between $\gamma_{B}^w$ and $\gamma_{B}^s$, we also computed the errors in $\frac{d e_B}{d\gamma_{B}}$ between the numerical and analytical results of the energy density function. In particular, we found that the absolute percentage error is given by:\\
$\Delta_{\%}(\frac{d e_{B}}{d \gamma_{B}}) \simeq 1.25 \cdot10^{-7}$, for  $\gamma_{B} = \gamma_{B}^w$, and:\\
$\Delta_{\%}(\frac{d e_{B}}{d \gamma_{B}}) \simeq 1.40 \cdot10^{-3}$, for $\gamma_{B} = \gamma_{B}^s$.   

In the intermediate region of couplings, \textit{i.e.} for $\gamma_B \in \left[\gamma_B^w, \gamma_B^s\right]$, we will build our Hermite interpolation scheme using Chebyshev polynomials as we did previously in the example of $\gamma_B \in[0.1,100]$. First of all we introduce the variable $\xi$ which ranges in $[-1,1]$ and is related to $\gamma_B$ as stated in Eq. (\ref{csi}). From this equation, fixing $\xi=-1$ for $\gamma_B=\gamma_B^w$ and $\xi=1$ for $\gamma_B = \gamma_B^s$, we find that the parameters $A$ and $B$ must be equal to: $A= \frac{134363}{100967}$, and $B= \frac{101433}{100967}$.

The mapping between $\xi$ and $\gamma_B$ is now completely defined and we can write the interpolation formula for the dimensionless ground state energy as:
\begin{equation}\label{Cheby_interp_egamma}
e_{B}^{i}(\xi) = \sum_{k=0}^{M+1} C_{k} \xi^{k}\,,
\end{equation} 
where the superscript $i$ stands for ``$interpolation$". While $M$ coefficients $C_k$ are determined by fixing the values of $e_B(\xi)$ at the Chebyshev points (\ref{Chebyshev-points}) to be the numerical resulting values [which come from solving Eq. (\ref{egammaB2}) using the interpolation scheme results for the density of pseudorapidities], the other two coefficients are fixed by requiring a continuity of the first order derivative of $e_B(\gamma_B)$ at the limiting region points $\gamma_B^w$ and $\gamma_B^s$, making such an interpolation one of the Hermite kind. 

Therefore, finally we can write a complete expression for the dimensionless ground state energy density function for the LL model as:
\begin{equation}
e_{B}^{3r}(\gamma_{B}) = \begin{cases}
e_{B}^{w}(\gamma_{B}) & \gamma_{B} < \gamma_{B}^w \\
e_{B}^{i}( \gamma_{B}) &  \gamma_{B} \in [ \gamma_{B}^w, \gamma_{B}^s] \\
e_{B}^{s}( \gamma_{B}) &  \gamma_{B} >  \gamma_{B}^s
\end{cases}\,, \label{three_range_egamma}
\end{equation}
where the superscript $3r$ stands for ``$\textit{three-ranges}$".

The remarkable feature of this expression is that it provides an analytical estimate of $e_{B}( \gamma_{B})$ for all values of $ \gamma_{B}$, allowing the ground state energy density to be evaluated very efficiently
at any given coupling. We refer to Appendix \ref{appendixA} for the values of the coefficients of the interpolating polynomial for the densities of pseudorapidities in Eq. (\ref{interp_ftilde}) used for the calculation of $e_{B}( \gamma_{B})$ at the selected values of $\xi_{i}( \gamma_{B})$, while we refer to Appendix \ref{appendixB} for the coefficients $C_{k}$ of Eq. (\ref{Cheby_interp_egamma}).

\begin{figure}[h!!]
\includegraphics[width=0.7\columnwidth]{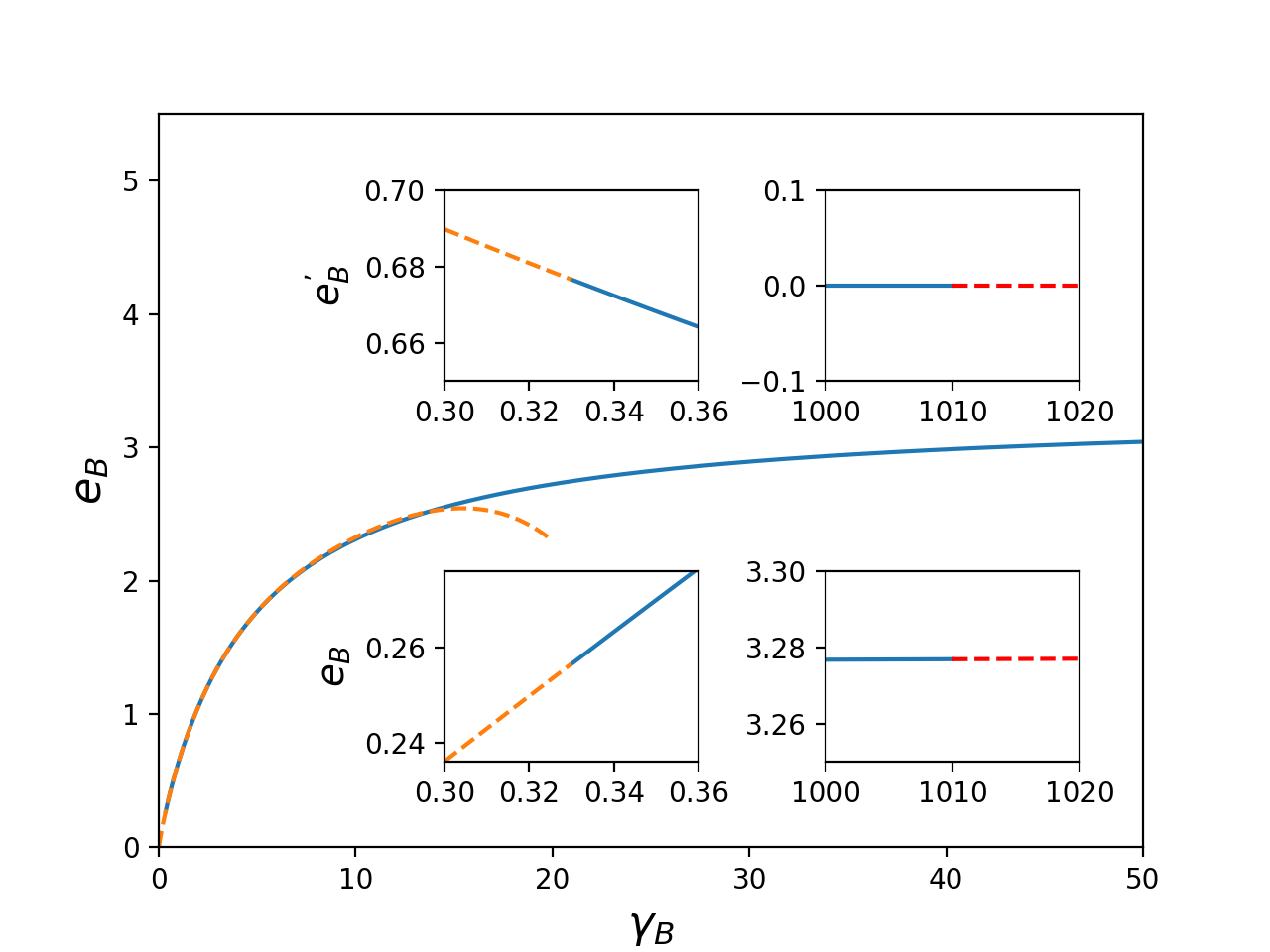}
\caption{Dimensionless ground state energy density function $e_{B}^{3r}(\gamma_{B})$ using the ``three-ranges" approach. The blue solid line refers to the interpolation scheme results, while the orange dashed line to $e_{B}^{w}$, \textit{i.e.} the power-series in $\gamma_{B}$ valid for low values of coupling strengths, and finally the red dashed line to $e_{B}^{s}$, \textit{i.e.} the power-series in $1/\gamma_{B}$ valid for strong interactions. In the main plot we can see that $e_{B}^{w}$ becomes less reliable as $\gamma_{B}$ increases, as we expect. In the lower-left inset and upper-left inset, we plot, respectively, the interpolation result $e_{B}^{i}$ and $e_{B}^{w}$  and their derivatives around $\gamma_{B} = \gamma_{B}^w$. Similarly, in the lower-right and upper-right inset, we plot, respectively, the interpolation result $e_{B}^{i}$ and $e_{B}^{s}$ and their derivatives around $\gamma_{B} = \gamma_{B}^s $.}
\label{fig8}
\end{figure}

In Fig. \ref{fig8} we plot the ground state energy density function of Eq. (\ref{three_range_egamma}) using $M=17$, for certain ranges of couplings. The blue solid line is the $e_B^i(\gamma_B)$ function, while the orange and red dashed lines represent the analytical behaviour at weak (\ref{low_gamma_Marino}) and strong (\ref{high_gamma_e}) couplings, respectively. The insets show the matching of the functions and their first order derivatives around the limiting regions points.

\begin{figure}[h!!]
\includegraphics[width=0.7\columnwidth]{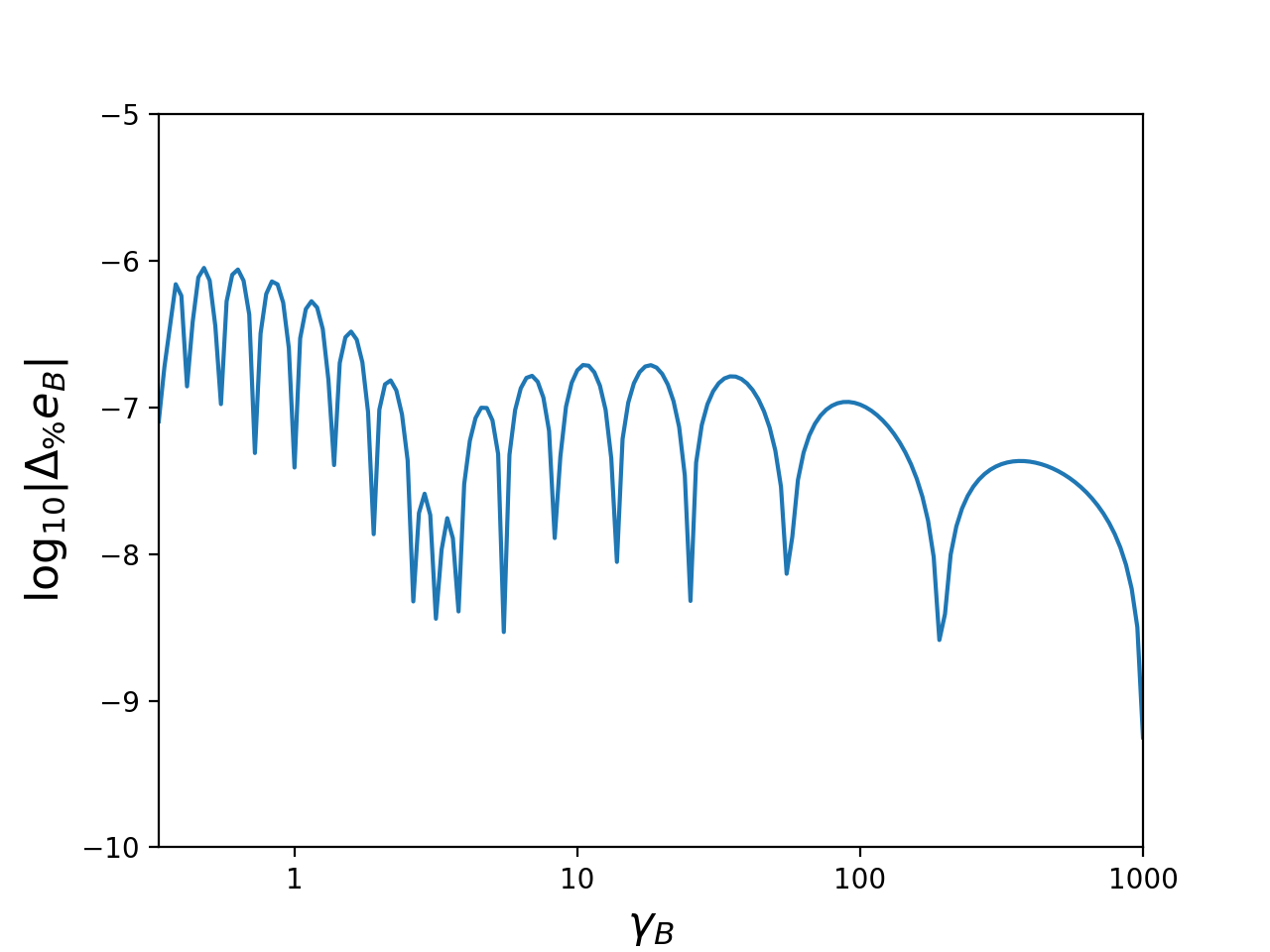}
\caption{Logarithm of the absolute value of the percentage error of $e_{B}^i$ \textit{vs} the dimensionless coupling, $\gamma_{B} \in[\gamma_B^w,\gamma_B^s]$, which is reported in log scale. These are the results coming from using $17 (+2)$ points for the Hermite--Chebyshev interpolation.}
\label{fig9}
\end{figure}

Finally, in Fig. \ref{fig9}, we plotted the relative percentage errors of this procedure from using only $17 (+2)$ interpolation points. The error of the interpolation is computed with respect to the numerical results coming from directly solving Eq. (\ref{egammaB2}), together with Eqs. (\ref{fBtilde}) and (\ref{norm_eq}). Note that we report the relative percentage error behaviour only in the region: $\gamma_B \in [\gamma_B^w, \gamma_B^s]$, since outside these regions we rely on the analytical forms of the dimensionless ground state energy density, which we have shown to describe the numerical results well.

Notice that the errors are remarkably small, even though we are using only $M=17$ for the interpolation scheme, and the error reaches a maximum of about $10^{-6}\%$ at weak couplings. The ``three-ranges" interpolation procedure thus performs considerably better than the simple single-range scheme of the previous section, for the case $\gamma_B\in[0.1,100]$. Given these very small errors, we may confidently use Eq. (\ref{three_range_egamma}) to compute interesting physical properties of the LL model. We will focus on the local two-particle correlation function and on the black solitons behaviour of the system, but other quantities such as the chemical potential, the pressure, the sound velocity and the Luttinger liquid parameter \cite{PethickBook}, could be obtained in the same way.

\section{Local two-particle correlation function for the Lieb--Liniger model}
\label{g_2_section}
\begin{figure}[h!!]
\includegraphics[width=0.7\columnwidth]{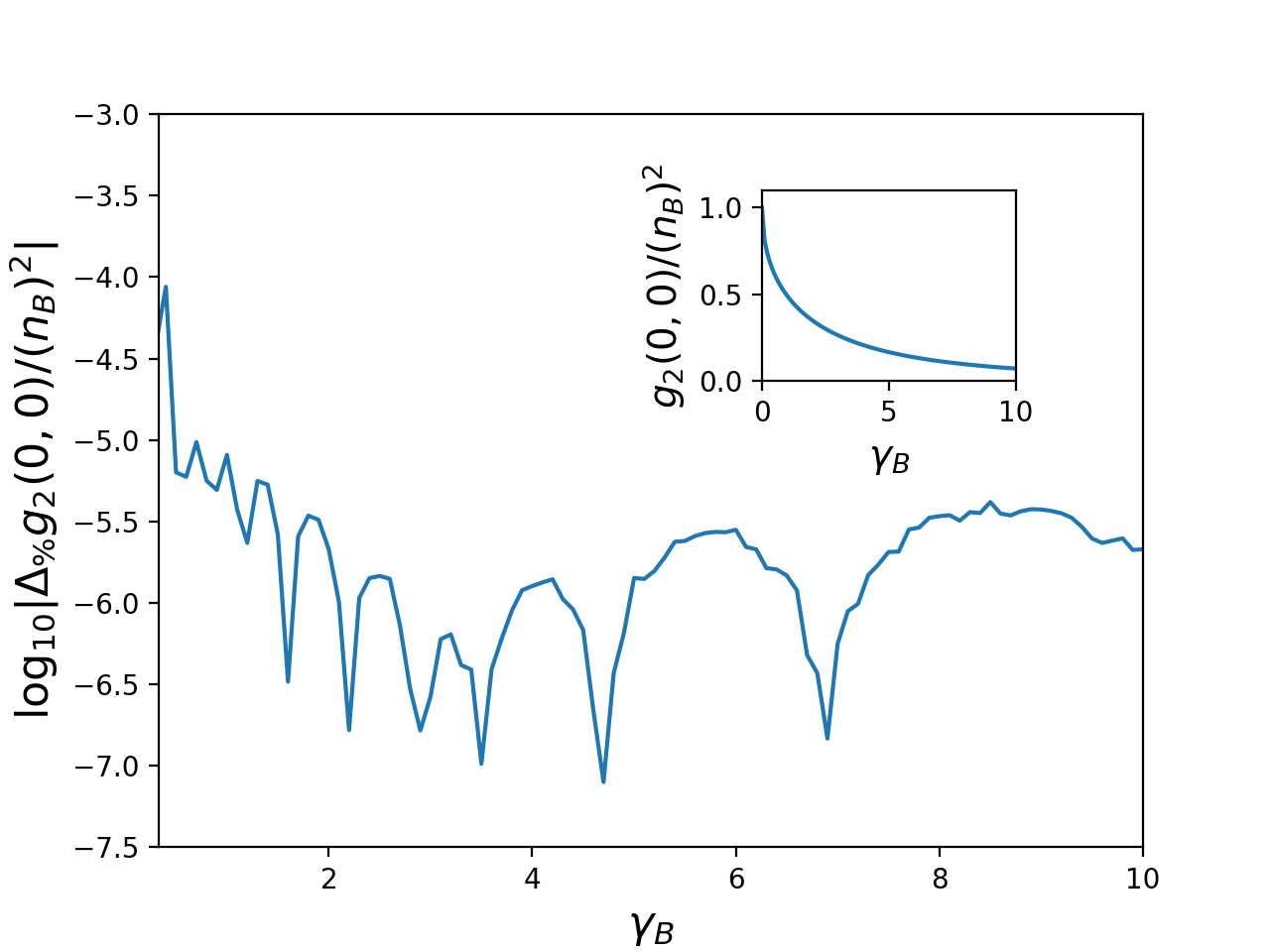}
\caption{Logarithm of the absolute value of the relative percentage error of the computation of the dimensionless diagonal local two-particle correlation function for the LL model, with respect to the dimensionless coupling, $\gamma_{B}$. The error is evaluated from comparing the results for $g_2(0,0)/(n_B)^2$ from using $e^{3r}(\gamma_B)$ with the numerical exact outcomes. In the inset there is shown the behaviour of the function $g_2(0,0)/(n_B)^2$ with respect to the dimensionless coupling, $\gamma_{B}$, obtained from using our interpolation scheme for the ground state energy density function. }
\label{fig11}
\end{figure}

The local two-particle correlation function is related to the dimensionless ground state energy density $e_B(\gamma_B)$.
The local two-particle correlation function $g_{2}$ is defined as:
\begin{equation}
g_{2}(x,x') = \langle \Psi^{\dagger}(x)\Psi(x)\Psi^\dagger (x')\Psi(x') \rangle\,,
\end{equation}
where $\Psi(x)$ is the bosonic matter-field in second quantization formalism. Using the Hellmann--Feynman theorem, one can prove that there exists a relation between the diagonal part of the local two-particle correlation function, $g_2(0,0)$, and the dimensionless ground state energy density, which reads \cite{Olshanii2003,PethickBook,Kormos2010}:
\begin{equation}
\frac{g_{2}(0,0)}{(n_B)^2} = \frac{de_{B}(\gamma_{B})}{d \gamma_{B}}\,.
\end{equation}
The function $g_{2}(0,0)$ is then obtained by differentiating $e_{B}^{3r}( \gamma_{B})$ of Eq. (\ref{three_range_egamma}) with respect to $\gamma_B$.
In the inset of Fig. \ref{fig11} we report the behaviour of $g_2(0,0)/(n_B)^2$ evaluated in this way, for different couplings. In the main plot we report the relative percentage error results coming from comparing the outcomes of $g_{2}(0,0)$ using $e^{3r}(\gamma_B)$ with the numerical exact one [\textit{i.e.} the results obtained from calculating directly the ground state energy density from Eq. (\ref{egammaB2}) and evaluating its first order derivative]. The maximum relative percentage error is below $10^{-4}\%$.

\section{Black solitons for the Lieb--Liniger model}
\label{solitons_section}
We now turn to a further property of the Lieb–Liniger gas whose determination requires knowledge of the dimensionless ground state energy density at all couplings: Black-soliton solutions.
In order to do so we consider the lowest order approximation in the Bogoliubov prescription, which corresponds to replacing the quantum fields $\Psi(x)$ in Eq. (\ref{H_LL}) with a classical field $\psi(x)$, which corresponds to its mean value. In the usual mean-field approach one has the Hamiltonian in which the interaction energy is proportional to $|\psi|^4$, whose equation of motion is the usual
Gross-Pitaevskii equation (or non-linear Schr\"odinger equation) $i \hbar \frac{\partial \psi}{\partial t}=-\frac{\hbar^2}{2m} \frac{\partial^2}{\partial x^2} \psi+c_B |\psi|^2 \psi$. 

An improved mean-field approximation can be obtained by
taking advantage of the analytical solution for the ground state energy
of the LL model (\ref{H_LL}). The approximated Hamiltonian then reads:
\begin{equation}
\label{H_LL_2}
H_B=\int_{0}^{L}\,dx\, \left\{ \frac{\hbar^2}{2m_B} \frac{\partial \psi^{\dagger}}{\partial x} \frac{\partial \psi }{\partial x} + G[n_B]\right\}\,,
\end{equation} 
where $G[n_{B}]$ is the interaction term, which depends on the density of the gas $n_{B}$. If the gas is homogeneous, we have that:
\begin{equation}
E = H_{B}[\rho] = L\,G[n_{B}]\,,
\end{equation}
that is, the interaction term corresponds to the energy per unit length of the system. Now, taking the thermodynamic limit, since the total energy of the system can be written as 
\begin{equation}\nonumber
E =\frac{\hbar^2}{2m_B}\, n_{B}^{3}\,L\, e_{B}(\gamma_{B})\,,
\end{equation}
this implies that:
\begin{equation}
G[n_{B}] =\frac{\hbar^2}{2m_B}\, n_{B}^{3}\,e_{B}(\gamma_{B}) \,.
\end{equation}
We recall that also $\gamma_B$ depends on the density of the gas according to Eq. (\ref{def_gamma_B}). 

The equation of motion in this improved mean field approach is then a generalization of the time-dependent Gross--Pitaevskii equation,
which reads \cite{Dunjko2001,Schmidt2006,Peotta2014,Choi2015}:
\begin{equation}
i\hbar \frac{\partial}{\partial t} \psi(x,t) =- \frac{\hbar^2}{2m_B} \frac{\partial^2 }{\partial x^2} \psi(x,t) + f\left[n_B\right] \,\psi(x,t)\,,
\end{equation}
referred to as the generalized Gross--Pitaevskii equation (or generalized
non-linear Schr\"odinger equation), where $\psi(x,t)$ is the wavefunction describing a generic state of the system. The
generalized Gross--Pitaevskii equation is derived from taking the functional derivative of the Hamiltonian $H_B$ of Eq. (\ref{H_LL_2}), and therefore the functional $f[n_{B}]$ is related to the functional $G[n_{B}]$ by:
\begin{equation}
f[n_{B}] = \frac{\delta G}{\delta n_{B}} = \frac{\hbar^2}{2m_B} \,\left[ 3\,e_{B}(\gamma_{B}) - \gamma_{B}\frac{d e}{d \gamma_{B}}\right]n_{B}^{2} \, .
\end{equation}

If we focus on the case $\psi(x,t) = e^{-i\frac{t}{\hbar}\mu}\, \psi(x)$, where $\mu$ is the chemical potential \cite{PitaevskiiBook},  
then the time-independent generalized Gross--Pitaevskii equation reads:
\begin{equation}\label{GGPE}
\mu\psi(x) = - \frac{\hbar^2}{2m_B} \frac{d^{2}}{dx^{2}}\psi(x) + f[n_{B}]\psi(x)\,.
\end{equation}
Since we will be interested in black soliton solutions, we stress that, although $\psi(x)$ describes an inhomogeneous soliton profile, the underlying system is homogeneous with density $n_B = N_B/L$. Because of the inhomogeneity introduced by the soliton structure, the gas density will depend on the position variable. 

We now seek a solution of Eq. (\ref{GGPE}) that corresponds to a black soliton, which is characterized by a localized drop of the density. We will take $x=0$ as the point where the soliton minimum occurs, and we denote by $n_{B,\infty}$ the density at infinity, which corresponds to the bulk density, providing a convenient scale for defining dimensionless variables. Moreover, black soliton solutions require that the derivatives of $\psi(x)$ vanish at infinity and that $\psi(x)$ be monotonically increasing \cite{PitaevskiiBook}. 
With this choice of boundary conditions, we have that Eq. (\ref{GGPE}) implies:
\begin{equation}
\mu =  f[n_{B,\infty}]\,,
\end{equation}
at infinite distances. We define the healing length $\xi_H$ as the scale at which the kinetic contribution becomes comparable to the interaction term. Therefore:
\begin{equation}
\frac{\hbar^2}{2m_B \xi_H^{2}} = f[n_{B, \infty}] \quad \rightarrow \quad \xi_H = \frac{1}{\sqrt{\frac{2m_B}{\hbar^2} f[n_{B,\infty}]}} = \frac{1}{\sqrt{\frac{2m_B}{\hbar^2}\mu}} \,,
\end{equation}
which relates the two length scales of the system: $n_{B,\infty}$ and $\xi_H$. This allows us to define both a dimensionless length: $\tilde{x} = \frac{x}{\xi_H}$, and a dimensionless wavefunction: $\varphi(\tilde{x}) = \psi(\xi_H \tilde{x})/\sqrt{n_{B,\infty}}$. With these notations, Eq. (\ref{GGPE}) can be rewritten in dimensionless units as:
\begin{equation}\label{AGGPE}
\varphi(\tilde{x}) = - \frac{d^{2}}{d\tilde{x}^{2}}\varphi(\tilde{x}) + h[\varphi^{2}]\varphi(\tilde{x}) \,,
\end{equation}
where we have used the fact that: $n_B = \left | \psi \right |^2 = n_{B,\infty}\left |\varphi\right|^2$, and we have defined:
\begin{equation}
h[\varphi^{2}] = \frac{2m_B \xi_H^{2}}{\hbar^2}\, f[n_{B,\infty}\varphi^{2}]\,.
\end{equation}
Notice that, since $\varphi(\tilde{x}\rightarrow\infty)=1$, then $h[1]=1$. In analogy with the functional $G[n_B]$, we now define a functional $H[\varphi^{2}]$ related to $h[\varphi^2]$ as:
\begin{equation}
h[\varphi^{2}] = \frac{\delta}{\delta \varphi^{2}}H[\varphi^{2}] \,.
\end{equation}
This functional form, allows to write Eq. (\ref{AGGPE}) in a form reminiscent of Newton's second law:
\begin{equation}\label{modNew}
\varphi^{''}(\tilde{x}) = -\frac{\delta}{\delta \varphi} V[\varphi]\,,
\end{equation}
where with the prime we have indicated the derivative with respect to $\tilde{x}$, and the (fictitious) `potential' $V[\varphi]$ is written as:
\begin{equation}\label{potential}
V[\varphi] = \frac{1}{2}\varphi^{2} - \frac{1}{2}H[\varphi^{2}]\,.
\end{equation}
In analogy with Newton's equation, we find a constant of motion, $\tilde{E}$, given by:
\begin{equation}\label{modenergy}
\tilde{E} = \frac{1}{2}\left(\varphi^{'}\right)^2 +  V[\varphi]\,,
\end{equation}
indeed, differentiating $\tilde{E}$ with respect to $\tilde{x}$, gives: 
\begin{equation}\nonumber
\tilde{E}^{'} = \varphi^{'}\left[ \varphi^{''} +  \frac{\delta}{\delta \varphi} V[\varphi]\right] = 0\,,
\end{equation}
where Eq. (\ref{modNew}) has been used. Since $\tilde{E}$ is a constant of motion, it is convenient to evaluate it at infinity. For a black soliton the derivatives of the wavefunction vanish there, so that:
\begin{equation}\label{E_fixed}
\tilde{E} =V[1] =  \frac{1}{2} - \frac{1}{2}H[1] \, .
\end{equation}

A black-soliton solution can now be constructed. From Eq. (\ref{modenergy}) we can write:
\begin{equation}
\varphi'(\tilde{x}) = \sqrt{2(\tilde{E} - V[\varphi])}\,,
\end{equation}
which in turn gives:
\begin{equation}
\tilde{x} = \int_{0}^{\varphi(\tilde{x})}  \frac{d\bar{\varphi}}{\sqrt{2(\tilde{E} - V[\bar{\varphi}])}}\,.
\end{equation}
This last expression can also be rewritten as:
\begin{equation}\label{xvsphi}
\tilde{x} =  \int_{0}^{\varphi(\tilde{x})}  \frac{d\bar{\varphi}}{\sqrt{1 - \bar{\varphi}^{2} - H[1] + H[\bar{\varphi}^{2}]}} \, ,
\end{equation}
by using Eqs. (\ref{potential}) and (\ref{E_fixed}).
We are then led to solving Eq. (\ref{xvsphi}).

We note that this procedure, until now, is completely general and does not depend on a particular non-linearity. In the case of the LL model, from the above equations, we have that:
\begin{equation}
H[\varphi^{2}] = \frac{e\left( \frac{c_{B}}{n_{B,\infty}\varphi^{2}} \right)\varphi^{6}}{3e\left(\frac{c_{B}}{n_{B,\infty}} \right) - \frac{c_{B}}{n_{B,\infty}}\frac{d e}{d \gamma_{B,\infty}}\Big|_{\gamma_{B,\infty} = \frac{c_{B}}{n_{B,\infty} }}}\,.
\end{equation}

Eq. (\ref{xvsphi}) can therefore be solved numerically by determining the value of $\tilde{x}$ corresponding to each $\varphi \in [0,1]$, exploiting the monotonicity of the wavefunction. To this end we make the following substitution in the integral:
\begin{equation}\nonumber
t = 2\frac{\bar{\varphi}}{\varphi(\tilde{x})} - 1 
\end{equation} 
which gives:
\begin{equation}\label{xvsphi2}
\tilde{x} =  \frac{1}{2} \int_{-1}^{1}  \frac{\varphi(\tilde{x})}{\sqrt{1 - \bar{\varphi}(t)^{2} - H[1] + H[\bar{\varphi}(t)^{2}]}}\,dt \, ,
\end{equation}
where:
\begin{equation}\nonumber
\bar{\varphi}(t) =  \frac{1}{2}(t+1)\varphi(\tilde{x}) \, .
\end{equation}

\begin{figure}[h!!]
\centering
\includegraphics[width=0.7\columnwidth]{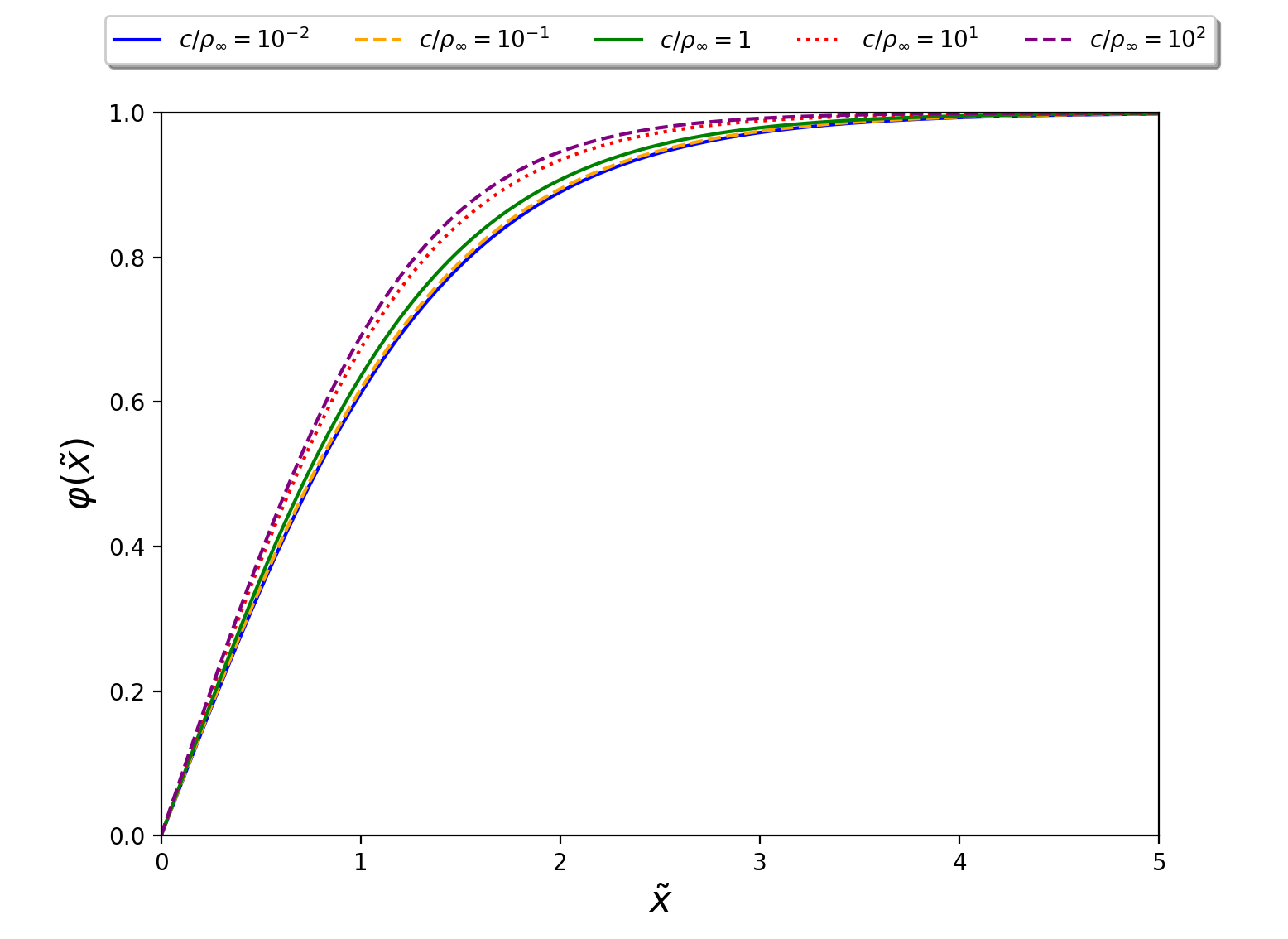}
\caption{Black soliton dimensionless wavefunction solutions of the generalized
  Gross-Pitaevskii equation, obtained by numerically solving Eq. (\ref{xvsphi2}), for different values of the ratio $c_{B}/n_{B,\infty}$.}
\label{fig12}
\end{figure}

\begin{figure}[h!!]
\centering
\includegraphics[width=0.7\columnwidth]{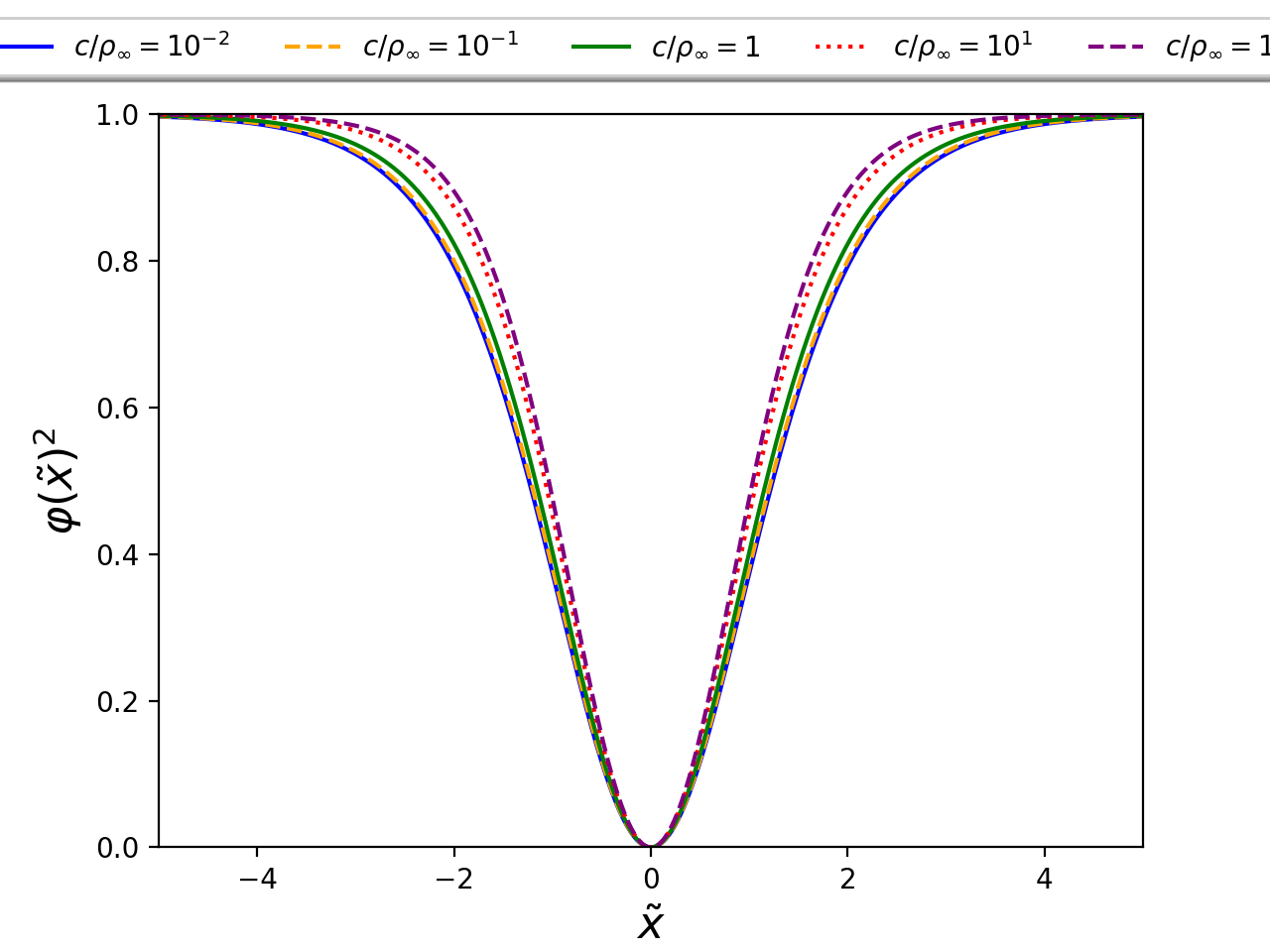}
\caption{Black soliton solutions densities of the generalized Gross-Pitaevskii  equation, obtained by numerically solving Eq. (\ref{xvsphi2}), for different values of the ratio $c_{B}/n_{B,\infty}$.}
\label{fig13}
\end{figure}

We solve this equation numerically using Chebyshev quadrature with $100$ points, together with our ``three-ranges" expression of Eq. (\ref{three_range_egamma}) for the dimensionless ground state energy density, which therefore turns out to be extremely useful also in this context. In Fig. \ref{fig12}, we report the dimensionless wavefunction obtained by solving Eq. (\ref{xvsphi2}), for different values of the ratio $c_B/n_{B,\infty}$, while in Fig. \ref{fig13} we plot the corresponding dimensionless density associated with the black soliton solutions.

\begin{figure}[h!!]
\centering
\includegraphics[width=0.7\columnwidth]{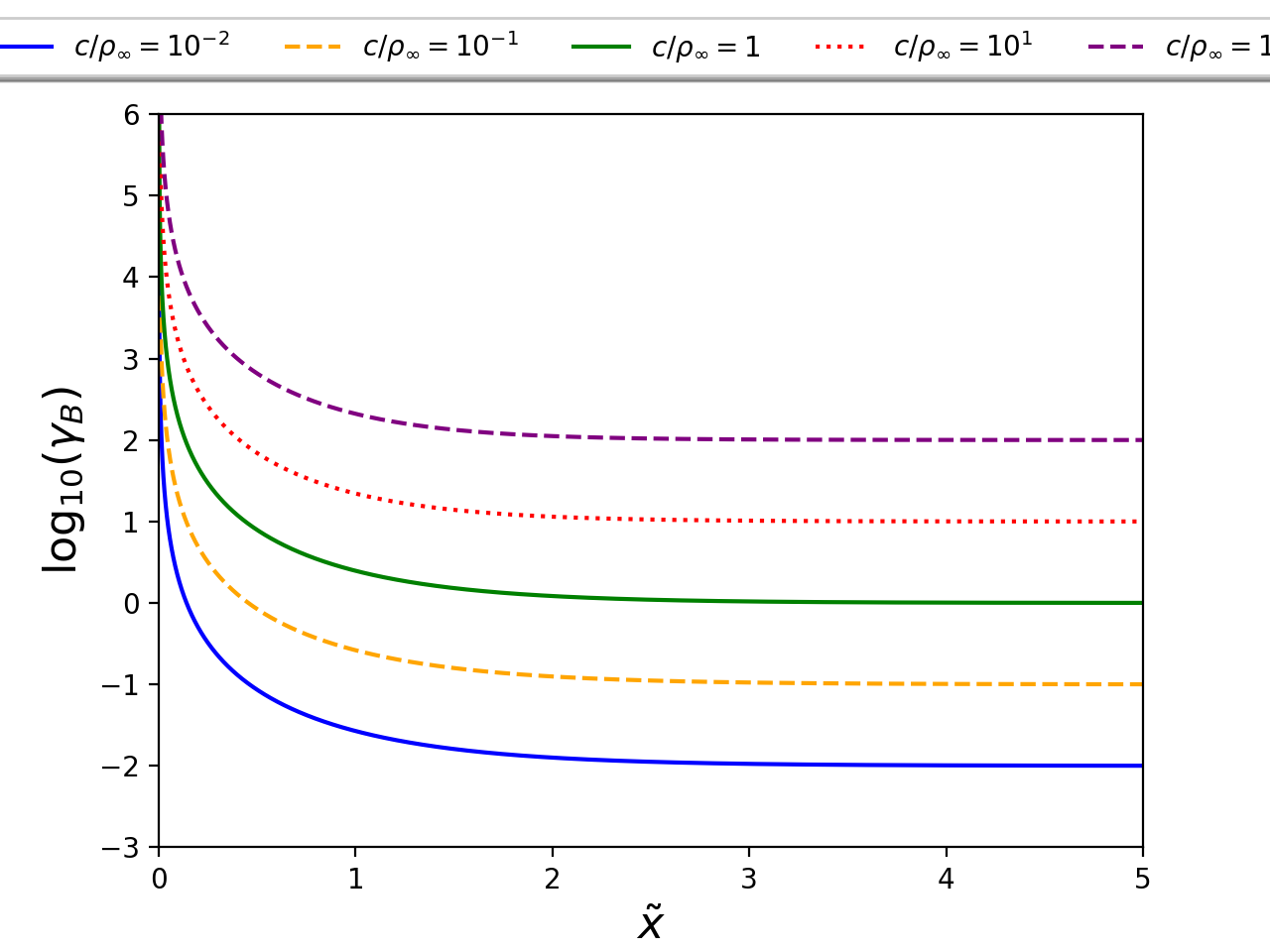}
\caption{Local dimensionless LL interaction parameter $\gamma_{B}(\tilde{x})$
  {\it vs} the dimensionless position variable $\tilde{x} = x/\xi_H$, for different values of the ratio $c_{B}/n_{B,\infty}$. Around $\tilde{x}=0$ the interaction parameter diverges, as it should since at $\tilde{x}=0$ the black soliton has its minimum, where the density vanishes.}
\label{fig14}
\end{figure}

We note that the density rises more steeply at larger $c_{B}/n_{B,\infty}$. This can be understood intuitively by considering that higher values of $c_{B}/n_{B,\infty}$ mean a stronger repulsive interaction away from the soliton minimum. This stronger repulsion competes with the tendency of particles to move away from $\tilde{x}=0$. To quantify this we can define a local interaction parameter
$\gamma_{B}(\tilde{x}) = c_{B}/[n_{B,\infty}\varphi^{2}(\tilde{x})]$, which diverges for $\tilde{x} \to 0$, as can be seen in Fig. \ref{fig14}. Finally notice that in the Tonks--Girardeau limit, $c_{B}/n_{B,\infty} \rightarrow \infty$, Eq. (\ref{xvsphi}) can be written as:
\begin{equation}
\frac{\tilde{x}}{\sqrt{3}} =\int_{0}^{\varphi(\tilde{x})} \frac{d\bar{\varphi}}{\sqrt{2 - 3\bar{\varphi}^{2} + \bar{\varphi}^{6} }} = \int_{0}^{\varphi(\tilde{x})} \frac{d\bar{\varphi}}{(1-\bar{\varphi})(1 + \bar{\varphi})\sqrt{2 + \bar{\varphi}^{2}}}\,,
\end{equation}
and can be solved analytically giving the dimensionless density:
\begin{equation}\label{blackTG}
\left[\varphi(\tilde{x})\right]^{2} = \frac{2\tanh\left(\tilde{x}\right)^{2}}{3 -\tanh\left(\tilde{x}\right)^{2} }\,,
\end{equation}
thus recovering the result of \cite{Kolomeisky2000}.
We note that the numerical result for $c_{B}/n_{B,\infty} = 100$ already approximates this limit closely.

In the opposite limit, \textit{i.e.} for $c_B/n_{B,\infty} \rightarrow 0$, one enters the Thomas--Fermi regime and recovers the black soliton solution of the Gross--Pitaevskii equation \cite{PitaevskiiBook}. In this case the black solitary wave density reads:
\begin{equation}
\left[\varphi(\tilde{x})\right]^2 = \tanh^{2}\left(\frac{\tilde{x}}{\sqrt{2}}\right)\,.
\end{equation}

\section{Conclusions}
\label{concl}
In this work, we provided a method for finding very accurate approximations for the normalized solutions of Love equations of both kinds,
for a wide range of the values of the interaction strength $\gamma$.
We used a Hermite interpolation with Chebyshev polynomials.
For the normalized solutions, we obtain an excellent analytical approximation, with maximum absolute value for the relative percentage error below $10^{-6}\%$ ($10^{-3}\%$) for the first (second) kind. The remarkable advantage over direct numerical solution is that the result is a simple analytical expression, using the coefficients given in Appendix \ref{appendixA} for $M=17$, which is very efficient for computational purposes.

We also used the same technique in order to find an analytical expression for the second, fourth and sixth momenta. Here the accumulation of error is more pronounced, and the estimated errors for the momenta range from about $10^{-2}\%$ to $3\%$. We then proceeded to find an analytical expression for the ground state dimensionless energy density of the Lieb-Liniger model, which is related to the second momentum of Love integral equation of the first kind. In order to find an analytical approximation valid for all values of the interaction parameter $\gamma$ and with a lower percentage error, we combined the known weak and strong coupling expansions of the energy density with the Hermite--Chebyshev interpolation technique at intermediate couplings, whose coefficients are given in
Appendix \ref{appendixB}. For intermediate values the maximum percentage error is about $10^{-6}\%$. We have therefore found an excellent approximation for the ground state energy density which has the great advantage of being a rather simple analytical expression, valid for all values of $\gamma$. To our knowledge, a single closed-form expression of this accuracy covering the entire range of $\gamma$ has not previously been tabulated.

As an application, we computed the dimensionless local two-particle correlation function in the origin. Again, the maximum percentage error for this quantity with respect to the numerical estimate is extremely low, of about $10^{-4}\%$.
Finally, we applied our expression for the dimensionless ground state energy to the black-soliton solutions of the Lieb–Liniger model, obtaining numerical density profiles. We stress the practical convenience of Eq. (\ref{three_range_egamma}) in this context: Evaluating the soliton profile requires $e_B(\gamma_B)$ and its derivative at a large number of points along the integration path, a task that would otherwise call for repeated numerical solution of the Bethe-ansatz equations.

Further applications, such as the study of the ground state solution of a dilute one--dimensional Bose gas in a trapping potential, or its non-equilibrium dynamics obtained by solving the non-linear Schr\"odinger equation as in \cite{Choi2015}, are of interest and will be addressed in future work.

\appendix

\section{Chebyshev coefficients for the dimensionless densities}\label{appendixA}

In this Appendix, we report, in Tables \ref{tab1}-\ref{tab4}, the coefficients of the interpolating polynomials of the normalized solutions of Love equation of first kind as written in Eq. (\ref{interp_ftilde}), for different values of $\gamma_{B}$. These were used to compute the dimensionless energy function $e_{B}(\gamma_{B})$ from Eq. (\ref{egammaB2}) and the resulting values to construct the interpolation formula as in Eq. (\ref{Cheby_interp_egamma}). The values for $\gamma_{B}$ presented in the table, are obtained by inverting Eq. (\ref{csi}), for the values of the auxiliary variable $\xi$ given by Eq. (\ref{Chebyshev-points}).

\begin{table}[H]
\centering
\caption{Chebyshev coefficients for the normalized solutions of Love equation of first kind for different values of the coupling strength.}
\label{tab1}
\hspace*{-0.9cm}
\begin{tabular} {|c|c|c|c|c|c|} 
\hline
 & $\gamma_{B} = 0.33$ & $\gamma_{B} = 0.352549754209764$ & $\gamma_{B} = 0.421968451186952$ & $\gamma_{B} = 0.543865810816466$ & $\gamma_{B} = 0.728686901148192$  \\
 \hline
$\tilde{C}_{B,0}$ & $0.567566935633021$ & $0.566117518286331$ & $0.562071868124157$ & $0.556111855989179$ &  $0.548922254752821$\\
$\tilde{C}_{B,1}$ & $-0.181087651541227$ &  $-0.178084326004653$ & $-0.169642307754557$ & $-0.156991674757648$ & $-0.141203615715813$\\
$\tilde{C}_{B,2}$ & $-0.0332918870849321$ & $-0.0321986416989924$ & $-0.0290272314367762$ & $-0.0239320542178306$ & $-0.0169393200032693$\\
$\tilde{C}_{B,3}$ & $-0.0104856563214514$ & $-0.00962681168152339$ & $-0.00701938536221686$ &  $-0.00263869095994641$ & $0.0028564369358436$\\
$\tilde{C}_{B,4}$ & $-0.00182570181761927$ & $-0.000934541773581208$ & $0.00167522200425051$ & $0.00530988734781651$ & $0.00756500818424641$\\
$\tilde{C}_{B,5}$ & $0.00336426813987679$ & $0.00396718751409075$ & $0.00576979497044618$ & $0.00701597101050628$ & $0.00481944948755777$\\
$\tilde{C}_{B,6}$ & $0.00185651150852466$ & $0.00476126778627769$ & $0.00739758716656256$ & $0.00452916056945378$ & $9.02153715742287\cdot10^{-5}$\\
$\tilde{C}_{B,7}$ & $0.0352500200519264$ & $0.0166277475645767$ & $-0.00273601504488668$ & $-0.00133286674726581$ & $-0.00143611404938058$\\
$\tilde{C}_{B,8}$ & $-0.158473395764371$ & $-0.0606721012876495$ & $0.0331978538407307$ & $0.00570488980982263$ & $-0.00443948636781177$\\
$\tilde{C}_{B,9}$ & $0.664514546733039$ & $0.272531276451783$ & $-0.121145777547971$ & $-0.0312844402972212$ & $0.0112937360196075$\\
$\tilde{C}_{B,10}$ & $-2.03443323814027$ & $-0.875561724413104$ & $0.319174097032559$ & $0.082305140383089$ & $-0.0314056947063606$\\
$\tilde{C}_{B,11}$ & $4.69716645856775$ & $2.10781131465778$ & $-0.649790164418862$ & $-0.180242303014874$ & $0.068403242756721$\\
$\tilde{C}_{B,12}$ & $-8.20327392181333$ & $-3.84998672022398$ & $0.969041486982181$ & $0.302095284619456$ & $-0.108555722527139$\\
$\tilde{C}_{B,13}$ & $10.7107121201319$ & $5.25437060764339$ & $-1.0520914166103$ & $-0.380880924146136$ & $0.129462725429476$\\
$\tilde{C}_{B,14}$ & $-10.2743126977068$ & $-5.27192897379479$ & $0.800163545256527$ & $0.35893798855224$ & $-0.113734116062226$\\
$\tilde{C}_{B,15}$ & $6.99136400049141$ & $3.75561988062242$ & $-0.392279440972518$ & $-0.241785340280486$ & $0.0704698091378746$\\
$\tilde{C}_{B,16}$ & $-3.17057989584824$ & $-1.78155697496749$ & $0.105873294233225$ & $0.108099167190419$ & $-0.0290143777596844$\\
$\tilde{C}_{B,17}$ & $0.853964800924773$ & $0.500669994684149$ & $-0.009226926729331$ & $-0.0284680531652042$ & $0.00713293619857366$\\
$\tilde{C}_{B,18}$ & $-0.102931082732883$ & $-0.0627633420730919$ & $-0.00111475383926687$ & $0.0033315150049834$ &  $-0.000794813366972264$\\
\hline
\end{tabular}
\end{table}

\begin{table}[H]
\centering
\caption{Chebyshev coefficients for the normalized solutions of Love equation of first kind for different values of the coupling strength.}
\label{tab2}
\begin{tabular} {|c|c|c|c|c|} 
\hline
 & $\gamma_{B} = 0.993716775409876$ & $\gamma_{B}= 1.36679726442637$ & $\gamma_{B}= 1.89324968961585$ & $\gamma_{B}= 2.64929559413603$  \\   
 \hline
$\tilde{C}_{B,0}$ & $0.541025627334755$ & $0.532821466947698$ & $0.524696020879854$  &  $0.517108322626197$\\
$\tilde{C}_{B,1}$ & $-0.122789590703395$ & $-0.101892757215025$ & $-0.0789852192256506$ &  $-0.0556191727699977$\\
$\tilde{C}_{B,2}$ & $-0.00841952165510538$ & $-5.18562578854457\cdot10^{-5}$ & $0.00542851551825972$  & $0.00648835577380876$\\
$\tilde{C}_{B,3}$ & $0.0071107886642332$ & $0.00745418280112161$ & $0.00436871712128594$ &  $0.00120541226140363$\\
$\tilde{C}_{B,4}$ & $0.00591935895022312$ & $0.00190632153416185$ &$-0.000488192890761522$  & $-0.000535683491514281$\\
$\tilde{C}_{B,5}$ & $0.000581100243794087$& $-0.00128188476889177$ & $-0.000579269753912311$ &$-3.79237778355927\cdot10^{-5}$ \\
$\tilde{C}_{B,6}$ & $-0.00132133274621803$& $-0.000249506137151389$ & $0.00086400665126021$ & $0.000285311083444821$\\
$\tilde{C}_{B,7}$ & $-0.00331637113698308$& $-0.00240027828718266$ & $-0.00384771347150574$  & $-0.00152315424955649$\\
$\tilde{C}_{B,8}$ & $0.0113031357612409$& $0.0122054040320497$ & $-0.00384771347150574$  & $0.00694463264386338$\\
$\tilde{C}_{B,9}$ & $-0.0372953591592855$& $-0.0409248038630977$&$0.0139193741515795$  & $-0.0234806590533766$\\
$\tilde{C}_{B,10}$ & $0.0963414833129976$& $0.104494725950281$ &$-0.0379581116746348$  & $0.05953446140697$\\
$\tilde{C}_{B,11}$ & $-0.186034195388452$& $-0.202277962420116$ &$0.0780178882759624$  & $-0.114236035745693$\\
$\tilde{C}_{B,12}$ & $0.272082826084284$& $0.296584482787163$ & $-0.12102867707345$ & $0.166115131488938$\\
$\tilde{C}_{B,13}$ & $-0.299880697609695$& $-0.327299179198888$ &$0.140755126134862$  & $-0.181841554267263$\\
$\tilde{C}_{B,14}$ & $0.244618349267389$& $0.267411475769833$ & $-0.1207004312116$ & $0.147387667132395$\\
$\tilde{C}_{B,15}$ & $-0.14309415902045$& $-0.156799040555825$ & $0.0739913294563581$ & $-0.0857507659754874$\\
$\tilde{C}_{B,16}$ & $0.056762013719023$& $0.0623743709315031$ & $-0.0306590647398981$  & $0.033855481551507$\\
$\tilde{C}_{B,17}$ & $-0.0136754496738198$& $-0.0150707545127731$ &$0.00768997202341284$  & $-0.00812119479728515$\\
$\tilde{C}_{B,18}$ & $0.00151068861480966$& $0.00166924034283699$ &$-0.000881419520727311$  & $0.000893322356674258$\\
\hline
\end{tabular} 
\end{table}

\begin{table}[H]
\centering
\caption{Chebyshev coefficients for the normalized solutions of Love equation of first kind for different values of the coupling strength.}
\label{tab3}
\begin{tabular} {|c|c|c|c|c|} 
\hline
 & $\gamma_{B}= 3.7697458847897$ & $\gamma_{B}= 5.51007229679089$ & $\gamma_{B}= 8.40152410630262$ & $\gamma_{B}= 13.6995595960527$  \\   
 \hline
$\tilde{C}_{B,0}$ & $0.510588500011774$ & $0.505614407917054$ & $0.5024028332559615$ & $0.500756095193617$ \\
$\tilde{C}_{B,1}$ & $-0.0344235074759821$& $-0.0180121378355927$ &$-0.0075780773833646$  &  $-0.00237961520765145$\\
$\tilde{C}_{B,2}$ & $0.0044318223436843$& $0.00194361959500424$ & $0.000536355971950372$ & $8.5052183092544\cdot10^{-5}$ \\
$\tilde{C}_{B,3}$ & $-6.44140806296441\cdot10^{-5}$& $-3.23555964836333\cdot10^{-5}$ & $3.58023373393375\cdot10^{-6}$ &$-2.91264421143953\cdot10^{-6}$  \\
$\tilde{C}_{B,4}$ & $-0.000116707064105585$& $2.27068029794054\cdot10^{-7}$ & $-0.000134504286300617$ &  $1.30450348379593\cdot10^{-6}$\\
$\tilde{C}_{B,5}$ & $4.1456292281214\cdot10^{-6}$& $-2.47090435916897\cdot10^{-5}$ &$-3.0305267451597\cdot10^{-5}$  & $-1.70783019133768\cdot10^{-5}$ \\
$\tilde{C}_{B,6}$ & $0.000195340493891223$& $0.00024945150497715$ &  $0.000281054200169888$ & $0.00015384841815173$ \\
$\tilde{C}_{B,7}$ & $-0.00125905689524705$& $-0.00159442705807315$  & $-0.00179374696456138$ & $-0.000948494395251196$ \\
$\tilde{C}_{B,8}$ & $0.00580692313271208$& $0.00729860844899388$  &$0.00821069597055493$  & $0.00417496991732077$ \\
$\tilde{C}_{B,9}$ & $-0.0197714442303559$& $-0.0246313839850254$  & $-0.0277101579445468$ & $-0.0134915740179207$ \\
$\tilde{C}_{B,10}$ & $0.0505126809081173$& $0.0623489197960692$ & $0.0701463196208615$ & $0.0325735356701119$ \\
$\tilde{C}_{B,11}$ & $-0.0976804407041794$& $-0.119433603613731$ & $-0.134381700008145$  & $-0.0593001260430194$ \\
$\tilde{C}_{B,12}$ & $0.143147668592303$& $0.173367567027433$ & $0.195086816898485$ & $0.0815345925932859$ \\
$\tilde{C}_{B,13}$ & $-0.157906171600792$& $-0.189440154149338$ & $-0.213198940924089$ & $-0.0841302210460655$ \\
$\tilde{C}_{B,14}$ & $0.12895242707621$& $0.153269883259225$ &  $0.17251522309877$ & $0.0640912734628448$ \\
$\tilde{C}_{B,15}$ & $-0.0755750125665294$&  $-0.0890135466839937$ & $-0.100204582987723$ & $-0.0349539540118729$ \\
$\tilde{C}_{B,16}$ & $0.0300492970812919$& $0.0350818286941494$ & $0.0394981572815896$ & $0.0129038841450243$\\
$\tilde{C}_{B,17}$ & $-0.00725725642921615$& $-0.00840091119304073$ & $-0.00945987922893601$ &  $-0.00288741938378695$\\
$\tilde{C}_{B,18}$ & $0.000803494867594064$& $0.000922550173563659$ & $0.00103899778758766$ &  $0.000295594531384081$\\
\hline
\end{tabular} 
\end{table}

\begin{table}[H]
\centering
\caption{Chebyshev coefficients for the normalized solutions of Love equation of first kind for different values of the coupling strength.}
\label{tab4}
\begin{tabular} {|c|c|c|c|c|} 
\hline
  & $\gamma_{B}= 24.9758490738671$  & $\gamma_{B}= 55.8522994485755$ &  $\gamma_{B} = 194.002365562876$ & $\gamma_{B} = 1010$\\ 
 \hline
$\tilde{C}_{B,0}$   & $0.500141618743294$ & $0.499996302872947$ & $0.4999798751831$ & $0.499979441751843$\\
$\tilde{C}_{B,1}$   & $-0.000490301809736882$ & $-5.06774298291383\cdot10^{-5}$ &  $-1.30991686260783\cdot10^{-6}$ & $-9.52832905542852\cdot10^{-9}$\\
$\tilde{C}_{B,2}$   & $6.39427536760411\cdot10^{-6}$ & $1.48612140742403\cdot10^{-7}$ & $1.55362114372944\cdot10^{-9}$ & $1.19794994929584\cdot10^{-9}$\\
$\tilde{C}_{B,3}$  & $-1.54128564972601\cdot10^{-7}$ & $-2.46711475238658\cdot10^{-8}$ & $-6.35373877157318\cdot10^{-8}$  & $-6.19867213818477\cdot10^{-8}$\\
$\tilde{C}_{B,4}$  & $1.80709051025535\cdot10^{-6}$ & $5.78921170283259\cdot10^{-7}$ &  $1.57824198250429\cdot10^{-6}$ & $1.52668155531094\cdot10^{-6}$\\
$\tilde{C}_{B,5}$  & $-2.60912223990792\cdot10^{-5}$ & $-7.8946885668425\cdot10^{-6}$ & $-2.26606862936175\cdot10^{-5}$ & $-2.17195140927471\cdot10^{-5}$ \\
$\tilde{C}_{B,6}$  & $0.00024269959317144$ & $6.83211847497165\cdot10^{-5}$ & $0.000209259950008987$&  $0.00019853224842508$\\
$\tilde{C}_{B,7}$  & $-0.00155209100859229$ & $-0.000400090294701315$ & $-0.00132703983133962$ &$-0.00124488898381797$ \\
$\tilde{C}_{B,8}$  & $0.00711906866099645$ &  $0.00165106450179274$ & $0.00602996810289655$ & $0.00558742631013616$\\
$\tilde{C}_{B,9}$  & $-0.0240759154925108$ & $-0.00492311314575602$ & $-0.020186233166454$ & $-0.0184581165206951$\\
$\tilde{C}_{B,10}$  & $0.0610747103539837$ & $0.0107472851924971$ & $0.0506585049336459$ &$0.045672242065529$ \\
$\tilde{C}_{B,11}$  & $-0.117250792842517$ & $-0.0172173736662214$ & $-0.0961702931787749$ & $-0.0854266312716667$\\
$\tilde{C}_{B,12}$ & $0.170577494927896$ & $0.0200433914373358$ & $0.138314386813007$ & $0.120978898122853$\\
$\tilde{C}_{B,13}$  & $-0.186804992696469$ & $-0.0164932094514618$ & $-0.149727915140516$ & $-0.128891569376129$\\
$\tilde{C}_{B,14}$  & $0.151469452044365$ & $0.00900989314177001$ &  $0.120007724600868$ & $0.101636454787784$\\
$\tilde{C}_{B,15}$  & $-0.0881572583952044$ & $-0.00275925816334754$ & $-0.0690490145028038$& $-0.0575177330584572$\\
$\tilde{C}_{B,16}$  & $0.0348172192673952$ & $0.000136030548787215$ &  $0.0269643674378442$ & $0.0220885078420937$\\
$\tilde{C}_{B,17}$ & $-0.00835448796445023$ & $0.000188567977277023$ & $-0.00639915871723547$ & $-0.00515462681817817$ \\
$\tilde{C}_{B,18}$  & $0.00091924965228061$ & $-4.41697292585832\cdot10^{-5}$ & $0.000696587470200982$ &$0.000551758280579244$ \\ 
\hline
\end{tabular} 
\end{table}

\section{Chebyshev coefficients for the dimensionless energy for the Lieb--Liniger model}\label{appendixB}

In this Appendix we report the coefficients of the interpolating polynomial $e_{B}^{i}(\gamma_{B})$ of Eq. (\ref{Cheby_interp_egamma}).\\

\begin{table}[H]
\centering
\caption{Chebyshev coefficients of the polynomial $e_{B}^i(\gamma_{B})$ for $\gamma_{B} \in [0.33, 1010]$}
\label{tab5}
\begin{tabular} {|c|c|} 
\hline
$C_{0}$ & $1.25564891735811$\\
$C_{1}$ & $1.3334354402797963$\\
$C_{2}$ & $0.4860354572947664$\\
$C_{3}$ & $-0.051984903874499507$\\
$C_{4}$ & $0.04904971501072371$\\
$C_{5}$ & $-0.001179939149676739$\\
$C_{6}$ & $-0.016280129473601624$\\
$C_{7}$ & $-0.013913030793665338$\\
$C_{8}$ & $-0.008138388523767335$\\
$C_{9}$ & $-0.0031840948669939976$\\
$C_{10}$ & $-0.0017175103845688048$\\
$C_{11}$ & $0.0002441458897381856$\\
$C_{12}$ & $0.0029410626430055585$\\
$C_{13}$ & $0.0014935402670359808$\\
$C_{14}$ & $-0.001372595919119918$\\
$C_{15}$ & $-0.0003179930299544679$\\
$C_{16}$ & $0.0007911621229777253$\\
$C_{17}$ & $-0.00003224993093157114$\\
$C_{18}$ & $-0.0002418397997916562$\\
\hline
\end{tabular} 
\end{table}

\end{document}